\documentclass[
twocolumn,
superscriptaddress,
pra,
longbibliography,
showpacs,
preprintnumbers,
amssymb,
floatfix
]{revtex4-1}

\usepackage{amsmath}
\usepackage{bm}
\usepackage{graphicx}
\usepackage[ansinew]{inputenc}
\usepackage{array}
\usepackage{color}
\usepackage{amstext}
\usepackage{amsthm} 
\usepackage{amssymb}
\usepackage{latexsym}
\usepackage{gensymb}
\usepackage{dsfont}
\usepackage{braket}
\usepackage[caption=false]{subfig}

\usepackage[makeroom]{cancel}

\usepackage{balance}
\usepackage{dcolumn}
\usepackage{bm}
\usepackage{bbm}
\usepackage{braket}
\usepackage{mathrsfs}
\usepackage{euscript}
\usepackage{comment}
\usepackage{txfonts}

\usepackage[dvipsnames]{xcolor}

\usepackage[
unicode=true,
bookmarks=false,
breaklinks=false,
pdfborder={0 0 1},
backref=false,
colorlinks=true,
citecolor=blue
]{hyperref}

\usepackage{tikz-cd}

\usepackage{mathrsfs}

\usepackage{amsthm}

\DeclareMathOperator{\Tr}{Tr}

\begin{document}


\title{Machine-Learning Prediction of Quantum Fisher Information from Collective Spin and Spectral Features
}

\author{Yusef Maleki}

 \affiliation{Institute for Quantum Science and Engineering, Texas A\&M University, College Station, Texas 77843, USA}

\author{Luis D. Zambrano Palma}
\affiliation{Institute for Quantum Science and Engineering, Texas A\&M University, College Station, Texas 77843, USA}

\date{\today}

\begin{abstract}
Quantum Fisher information (QFI) is a fundamental quantifier in quantum metrology, determining the ultimate precision achievable in parameter-estimation protocols through the quantum Cram\'er-Rao bound. However, direct evaluation of the QFI generally requires detailed knowledge of the density matrix, making it increasingly demanding as the Hilbert-space dimension grows. In this work, we investigate the extent to which the QFI of multipartite quantum systems can be predicted from a limited set of experimentally accessible quantities using support vector regression (SVR). By comparing different physically motivated features, we identify a dominant feature set governing QFI and show that the predictive power of collective spin moments alone decreases as system size and consequently Hilbert-space dimension grows. We demonstrate that QFI is governed primarily by the interplay between collective covariance and low-order spectral moments of the density matrix. 
Our results identify the physically relevant information sectors governing the QFI and demonstrate that accurate estimation of metrological sensitivity can be achieved from a restricted set of experimentally accessible quantities without requiring full quantum-state tomography.\end{abstract}

\maketitle

\section{Introduction}

Quantum technologies have rapidly evolved from fundamental theoretical concepts into practical platforms for information processing, enabling advances in quantum computing~\cite{ladd2010quantum,nielsen2010quantum}, quantum communication \cite{usenko2026continuous,hasan2023quantum}, quantum simulation \cite{halimeh2025cold,daley2022practical}, quantum illumination and radars~\cite{karsa2024quantum, palma2026optimal}, and quantum metrology~\cite{pirandola2018advances, huang2024entanglement}. Despite their diverse objectives, these applications share a common foundation: the encoding, manipulation, and extraction of information from quantum systems through controlled dynamical evolutions and measurement processes~\cite{vonNeumann1955,4h4b-3xss,doi:10.1142/S0219477525400280}.

Among these applications, quantum metrology has emerged as one of the most promising areas of quantum science, offering the possibility of surpassing classical precision limits through the exploitation of quantum resources~\cite{giovannetti2004quantum,giovannetti2006quantum,pezze2018quantum}. At the heart of quantum parameter-estimation theory lies the quantum Fisher information (QFI), a fundamental quantity that determines the ultimate precision attainable in the estimation of an unknown parameter through the quantum Cram\'er--Rao bound~\cite{paris2009quantum,helstrom1969quantum,gudder1985holevo}. In principle, the QFI quantifies the sensitivity of a quantum state under infinitesimal parameter variations and therefore provides a direct measure of its metrological usefulness. Consequently, states possessing large QFI are regarded as valuable resources for quantum-enhanced estimation and for approaching the Heisenberg limit, thereby enabling precision beyond the standard quantum limit~\cite{pezze2009entanglement}. Considering its fundamental connection with the distinguishability of neighboring quantum states, the QFI has also found important applications beyond quantum metrology, including the characterization of quantum phase transitions~\cite{hauke2016measuring}, quantum Zeno dynamics~\cite{smerzi2012zeno}, and quantum speed limits~\cite{taddei2013quantum,
maleki2024universal}.

Despite its central role, QFI is not generally a directly accessible experimental quantity. For mixed and multipartite states, its evaluation depends on the spectral structure of the density operator and often requires quantum-state tomography, whose cost increases rapidly with system size \cite{vidrighin2014joint}. This creates a practical gap between the fundamental importance of the QFI and its  experimental accessibility. In many quantum platforms, however, low-order collective observables, such as spin moments, variances, and symmetrized correlations, can be accessed with substantially lower experimental cost than the full density matrix. This motivates a natural question: to what extent is the QFI encoded in such a restricted set of collective observables?

Machine learning provides a natural framework for addressing this question because the relation between collective observables and QFI is generally nonlinear and state dependent. Rather than reconstructing the full density matrix, a trained regression model can learn the mapping from physically motivated observables to the corresponding QFI. 

Machine-learning methods have already been widely used across quantum science, including adaptive quantum estimation and quantum-state tomography~\cite{quek2021adaptive,lohani2020machine}, wave-function reconstruction~\cite{10.21468/SciPostPhys.7.1.009}, quantum error correction~\cite{nautrup2019optimizing,zeng2023approximate}, and the characterization of many-body quantum systems using physics-informed neural networks~\cite{ferrer2026physics}. In quantum computing~\cite{wise2021using,strikis2021learning,che2026quantum}, quantum communication~\cite{wallnofer2020machine,chin2021machine}, and quantum information science, machine learning has also become an important tool for detecting and characterizing quantum resources. Significant progress has been reported in entanglement detection and classification~\cite{varela2026entanglement,PhysRevLett.122.200401,luo2023detecting,PhysRevA.108.022427,khoo2021quantum,harney2020entanglement,lu2017separability,greenwood2023machine,martinez2026entanglement,vintskevich2023classification,gao2024correlation}, entanglement quantification~\cite{Feng_2024,lin2023quantifying,koutny2023deep}, EPR-steering classification and quantification~\cite{PhysRevLett.123.190401,wang2024deep,PhysRevA.104.052427,ren2019steerability,Zhang2020Machine,PhysRevA.105.032408}, quantum discord estimation~\cite{li2019machine,taghadomi2025effective,pan2025estimating}, and quantum coherence characterization~\cite{app14167312,lu2024quantum}. Machine-learning techniques have also been applied to quantum metrology and quantum-enhanced sensing protocols in recent years~\cite{proceedings2019012028,mantilla2026measurement,huang2025quantum,belliardo2024applications,fallani2022learning}. 

In this work, we investigate the extent to which the QFI is encoded in a restricted set of experimentally accessible quantities. Specifically, we ask whether the QFI can be accurately inferred from a reduced set of physically motivated features, including collective-spin observables and low-order spectral moments. Beyond numerical prediction, our goal is to identify which physical quantities carry the dominant information governing the QFI and how this information content changes with increasing system size.
Using large datasets of random quantum states composed of two to five qubits, we examine whether the nonlinear relation between these features and the QFI can be learned with high accuracy. This allows us to move beyond a purely predictive task and use the regression model as a diagnostic tool for the observable structure of metrological sensitivity. In particular, we analyze how QFI-relevant information is distributed among collective moments, correlation functions, and spectral descriptors of multipartite quantum states. 
This problem is motivated not only by practical applications in quantum metrology, but also by the mathematical structure of the QFI itself. Although direct evaluation of the QFI generally requires detailed knowledge of the density operator, including its spectral decomposition, our analysis suggests that a significant fraction of the relevant metrological information can be captured by a compact set of experimentally motivated quantities. This provides a route toward estimating metrological sensitivity without relying on complete quantum-state reconstruction.

To perform this task, we employ support vector regression (SVR), a supervised learning method well suited for nonlinear regression problems with physically structured input features. SVR has demonstrated strong performance in investigating a variety of physical systems while maintaining good generalization capability. Successful applications of support vector machine-based methods have been reported in nuclear and high-energy physics~\cite{yang2026alpha,whiteson2003support,jalili2024prediction,vitek2013towards}, condensed-matter and materials science~\cite{seko2014machine,gao2009accurate,balabin2011support,chowdhury2025exploring}, and the estimation of quantum coherence and entanglement measures~\cite{lin2026machine}. Inspired by these developments, we use SVR to learn the mapping from experimentally motivated quantities to the QFI and to assess how the measurable encoding of metrological information evolves as the Hilbert-space dimension increases.

This work is organized as follows. In Sec.~\ref{sec2}, we introduce the quantum Fisher information and the metrological framework considered in this study. In Section~\ref{sec3} we present the support vector regression methodology. In Sec.~\ref{sec4}, the relationship between the QFI and the selected features through a comparison of different kernel functions is investigated. In Section~\ref{sec5} we analyze the predictive power of collective-spin observables in two-qubit systems. In Sec.~\ref{sec6} we extend the analysis to multipartite quantum states and examines the role of low-order spectral moments in the prediction of the QFI. Finally, conclusions are presented in Sec.~\ref{sec8}.

\section{Quantum Fisher Information and Metrological Sensitivity} \label{sec2}

A central objective of quantum metrology is the estimation of an unknown parameter encoded in a quantum system with the highest achievable precision. In a general parameter-estimation protocol, an initial quantum state $\rho$ undergoes a parameter-dependent unitary transformation, $\rho(\theta)=e^{-i\theta \hat H}\rho\,e^{i\theta \hat H},$
where $\theta$ denotes the parameter of interest and $\hat H$ is the Hermitian generator responsible for the encoding process. Information about $\theta$ is extracted through a measurement described by a positive-operator-valued measure (POVM) $\{M_\mu\}$, producing outcomes with conditional probabilities $P(\mu|\theta)=\mathrm{Tr}\![\rho(\theta)M_\mu]$~\cite{maleki2022distributed}.
The sensitivity of the measurement outcomes to variations of the encoded parameter is quantified by the classical Fisher information~\cite{petz2011introduction,walborn2018quantum}
\begin{equation}
F_C=\sum_{\mu}
\frac{1}{P(\mu|\theta)}
\left[
\frac{\partial P(\mu|\theta)}
{\partial\theta}
\right]^2,
\end{equation}
which determines the precision of any unbiased estimator through the Cram\'er--Rao bound $\Delta\hat{\theta}\geq \frac{1}{\sqrt{mF_C}}$,
where $m$ is the number of independent experimental repetitions~\cite{helstrom1969quantum,gudder1985holevo}. Optimizing the Fisher information over all admissible quantum measurements defines the quantum Fisher information (QFI), $F_Q=\max_{\{M_\mu\}}F_C,$ which specifies the ultimate precision permitted by quantum mechanics through the quantum Cram\'er--Rao bound, $\Delta\hat{\theta}\geq \frac{1}{\sqrt{mF_Q}}$ \cite{maleki2021quantum2}.
Consequently, for a mixed quantum state with spectral decomposition $\rho=\sum_i \lambda_i |i\rangle\langle i|,$
the QFI associated with the generator $\hat H$ is given by~\cite{braunstein1994statistical}
\begin{equation}
F_Q[\rho,\hat H]
=
2\sum_{ij}
\frac{(\lambda_i-\lambda_j)^2}
{\lambda_i+\lambda_j}
|\langle i|\hat H|j\rangle|^2,
\label{qfi}
\end{equation}
where $\lambda_i$ and $|i\rangle$ denote the eigenvalues and eigenvectors of $\rho$, respectively.


Equation~(\ref{qfi}) shows that the QFI depends simultaneously on the coherence properties and spectral structure of the density operator through both eigenvalues and eigenvectors. As a consequence, its direct evaluation generally requires substantial information about the structure of quantum states, whose cost can increase rapidly with system size.

Throughout this work, we consider multi-qubit parameter-estimation protocols with the collective spin operators \cite{hyllus2012fisher}
\begin{equation}
\hat J_k=\frac{1}{2}\sum_{l=1}^{N}\sigma_k^{(l)},
\qquad
k=x,y,z,
\end{equation}
where $\sigma_k^{(l)}$ denotes the Pauli operator acting on the $l$th qubit. These operators represent natural observables in many quantum metrology and many-body sensing protocols. In our analysis, the QFI is evaluated with respect to the collective spin generator $\hat J_z$ \cite{maleki2021quantum}. Therefore, unless otherwise stated, all reported values of $F_Q$ correspond to $F_Q[\rho,\hat J_z]$.


\section{Machine Learning Framework} \label{sec3}

Among supervised machine-learning methods, support vector regression (SVR), derived from support vector machines (SVM), has emerged as a powerful tool in statistical learning theory \cite{vapnik2013nature,cortes1995support}. By constructing an optimal regression hyperplane through a constrained optimization procedure, SVR provides an efficient framework for high-dimensional feature spaces, limited datasets, and nonlinear regression problems. Unlike traditional SVM approaches designed for classification tasks, SVR aims to predict continuous quantities while maintaining the prediction error within a controlled tolerance margin. Its effectiveness becomes particularly evident when the input vectors are mapped into a higher-dimensional feature space through kernel functions, allowing nonlinear relationships in the original space to be treated in a computationally efficient manner.

Considering a dataset composed of $\{x_n,y_n\}$ for $n=1,\dots,N$, where $x_n$ denotes the input vector and $y_n$ represents the corresponding target value, the SVR model can be written as \cite{bishop2006pattern}
\begin{equation}
    f(x)=w^{T}\phi(x)+b,
\end{equation}
where $\phi(x)$ denotes the nonlinear mapping into the high-dimensional feature space, $w$ is the weight vector associated with the transformed features, and $b$ corresponds to the bias parameter. The optimization problem in SVR is formulated as
\begin{equation}
\begin{aligned}
    \min_{w,b,\xi,\xi^{*}}
    \quad &
    \frac{1}{2}\|w\|^{2}
    +
    C\sum_{n=1}^{N}(\xi_n+\xi_n^{*}) \\
    \textrm{s.t.} \quad &
    y_n-w^{T}\phi(x_n)-b \leq \epsilon+\xi_n, \\
    &
    w^{T}\phi(x_n)+b-y_n \leq \epsilon+\xi_n^{*}, \\
    &
    \xi_n,\xi_n^{*}\geq0,
\end{aligned}
\end{equation}
where the first term controls the model complexity through regularization, while the second term penalizes deviations larger than the prescribed tolerance margin. Here, $C$ is the regularization parameter controlling the trade-off between model flatness and training error, $\epsilon$ defines the tolerance region around the regression function, and $\xi_n$ and $\xi_n^{*}$ are slack variables introduced to account for deviations outside the tolerance region \cite{bishop2006pattern}.

SVR can be implemented using different kernel functions, including linear, polynomial, and radial basis function (RBF) kernels. The linear kernel is defined as
\begin{equation}
K(x_i,x_j)
=
\gamma x_i^{T}x_j,
\end{equation}
which corresponds to a linear mapping in the original feature space. Another commonly used choice is the polynomial kernel,
\begin{equation}
K(x_i,x_j)
=
(\gamma x_i^{T}x_j + c_0)^d,
\end{equation}
where $d$ denotes the polynomial degree, $\gamma$ controls the scaling of the inner product, and $c_0$ is a constant offset parameter.
However, among the most widely used kernels for nonlinear regression problems is the radial basis function (RBF) kernel, defined as
\begin{equation}
K(x_i,x_j)
=
\exp\left(
-\gamma \left\|x_i-x_j\right\|^{2}
\right),
\label{eq16}
\end{equation}
where $\gamma$ determines the characteristic scale of the kernel function and controls the influence of neighboring data points in the feature space. The predictive performance of the SVR model strongly depends on the proper selection of the hyperparameters $C$, $\epsilon$, and $\gamma$. In practice, these parameters are commonly optimized through a grid-search procedure combined with cross-validation, where different hyperparameter combinations are systematically explored until the desired predictive accuracy is achieved. For all calculations presented in this work, the SVR models were implemented using the \texttt{scikit-learn} machine-learning library~\cite{pedregosa2011scikit}.

The predictive performance of the regression model is quantified through the coefficient of determination $R^2$, the mean squared error (MSE), the root-mean-square error (RMSE), and the mean absolute error (MAE). These quantities provide complementary information regarding the accuracy, stability, and generalization capability of the regression model \cite{lin2026machine}.
The coefficient of determination is defined as
\begin{equation}
R^2
=
1-
\frac{
\sum_{n=1}^{N} \left[f(x_n)- y_n\right]^2
}{
\sum_{n=1}^{N} \left(y_n-\bar y\right)^2
},
\end{equation}
where $y_n$ and $f(x_n)$ denote the exact and predicted values, respectively, and $\bar y$ is the mean target value. The quantity $R^2$ measures the fraction of the variance of the target data explained by the regression model, with values closer to unity indicating higher predictive accuracy.
The mean squared error is given by $\mathrm{MSE}
=
\frac{1}{N}
\sum_{n=1}^{N}
\left[f(x_n)- y_n\right]^2$,
which quantifies the average quadratic deviation between the predicted and exact values, thereby assigning larger penalties to larger prediction errors. On the other hand,
the root-mean-square error is defined as $\mathrm{RMSE}
=
\sqrt{\mathrm{MSE}}$, providing an error measure with the same physical units as the target quantity and allowing a more direct interpretation of the typical prediction deviation. Furthermore, 
the mean absolute error is defined as $\mathrm{MAE}
=
\frac{1}{N}
\sum_{n=1}^{N}
\left|f(x_n)- y_n\right|$,
which measures the average absolute deviation between the predicted and exact values and is less sensitive to outliers than the MSE.

\section{Nonlinear Structure of Quantum Fisher Information in Collective Observables} \label{sec4}

We begin by investigating the informational structure underlying the QFI through the relation between collective observables and the corresponding metrological sensitivity. To this end, we constructed a broad ensemble of two-qubit quantum states containing pure, mixed, and hybrid density matrices, allowing the regression model to explore a large region of the two-qubit Hilbert space.

The training dataset consists of $2000$ random pure states, $6000$ random mixed states generated from convex combinations of random pure states, and $2000$ hybrid states interpolating between pure and diagonal density matrices. General two-qubit pure states were sampled according to $\ket{\psi}
=
a\ket{00}
+
b\ket{01}
+
c\ket{10}
+
d\ket{11},$
where the complex coefficients satisfy the normalization condition $|a|^2+|b|^2+|c|^2+|d|^2=1.$ And, the corresponding density operator is given by $\rho_{\mathrm{pure}}
=
\ket{\psi}\bra{\psi}$.
Moreover, the set of mixed states were generated as convex combinations of random pure states, $\rho_{\mathrm{mixed}}
=
\sum_{k=1}^{n}
p_k
\ket{\psi_k}\bra{\psi_k}$, with $\sum_{k=1}^{n} p_k = 1$, and $p_k \ge 0.$
To further increase the diversity of the state ensemble, we additionally considered hybrid pure-diagonal states of the form $\rho_{\mathrm{hybrid}}
=
t\,\rho_{\mathrm{diag}}
+
(1-t)\ket{\psi}\bra{\psi},$
where $0\le t\le1$ and $\rho_{\mathrm{diag}}
=
\mathrm{diag}
(\lambda_1,\lambda_2,\lambda_3,\lambda_4),$
with $\sum_{i=1}^{4}\lambda_i = 1,$ and $\lambda_i \ge 0.$
\begin{figure*}[!htbp]
\includegraphics[width=0.9\textwidth]{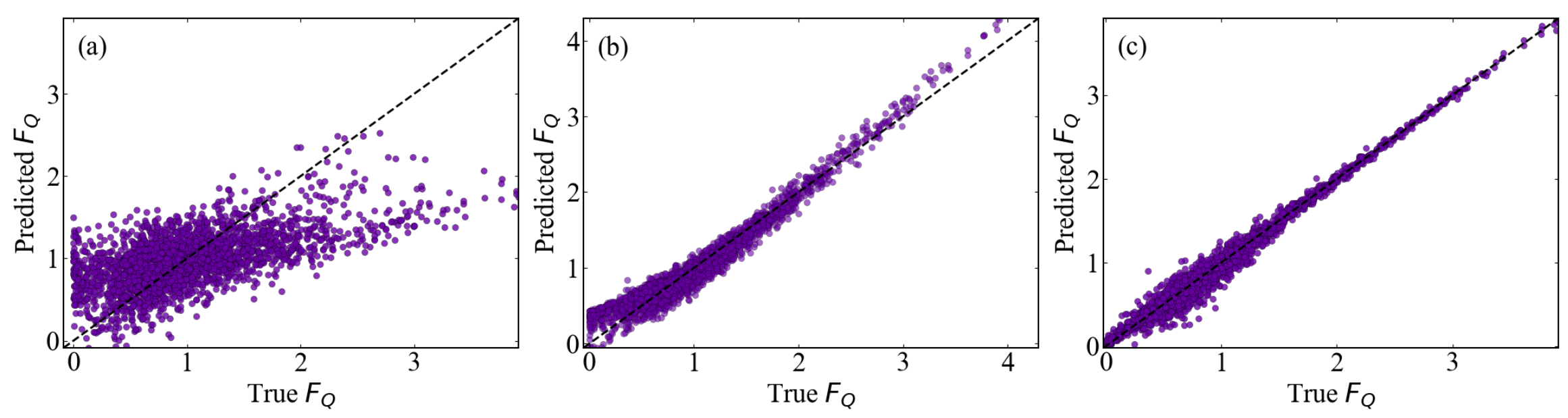} 
\caption{
Comparison between exact and SVR-predicted quantum Fisher information using (a) linear, (b) polynomial ($d=2$), and (c) RBF kernels. The dashed line denotes the ideal agreement between the exact and predicted QFI values.
}
\label{fig1}
\end{figure*}
The resulting ensemble therefore contains quantum states with broadly varying ranks, purities, coherence properties, and spectral structures, providing a nontrivial testbed for analyzing the observable-to-QFI mapping.

For all regression models, the complete dataset was randomly divided into training and test subsets, where $75\%$ of the states were used during the training stage and the remaining $25\%$ were reserved for independent performance evaluation. The test dataset was not employed during the optimization procedure, allowing an unbiased assessment of the generalization capability of the regression model.

The QFI is evaluated with respect to the collective generator $\hat J_z$, while the input vectors were constructed from collective spin moments and symmetrized correlators. In particular, the feature space is defined through the set~\cite{hyllus2012fisher,vitale2024robust},
\begin{equation}
\left\{
\langle \hat J_i\rangle,
\,
\langle \hat J_i\rangle^2,
\,
\langle \hat J_i^2\rangle,
\,
\left\langle
\hat J_i\hat J_j
+
\hat J_j\hat J_i
\right\rangle
\right\}, \label{setofkernel}
\end{equation}
where $\{i,j\}=x,y,z$.

The resulting kernel comparison is presented in Fig.~\ref{fig1}, where the predicted QFI values are plotted against the exact QFI values for the linear, polynomial, and RBF kernels. The dashed diagonal line shows the ideal predicted condition
\begin{equation}
F_Q^{\mathrm{predicted}}
=
F_Q^{\mathrm{exact}}.
\end{equation}
 As shown in Fig.~\ref{fig1}(a), the linear kernel exhibits substantial deviations from the ideal prediction line, particularly in the large-QFI regime. This behavior indicates that the relation between collective observables and the QFI cannot be accurately described through a purely linear mapping in feature space. Such a result is consistent with the nonlinear structure of the QFI itself, as manifested in Eq.~\eqref{qfi}, which depends nontrivially on both the spectral properties of the density operator and the transition matrix elements of the generator.

A substantial improvement is observed for the polynomial kernel shown in Fig.~\ref{fig1}(b), where the predicted values become significantly more concentrated around the ideal prediction line. This improvement reveals that nonlinear combinations of collective observables already encode an important fraction of the metrological information associated with the QFI. Nevertheless, noticeable deviations indicate that low-order polynomial nonlinearities are still insufficient to fully capture the complexity of the observable-to-QFI mapping.

The highest prediction accuracy is obtained with the RBF kernel [Fig.~\ref{fig1}(c)], where the predicted values become strongly localized around the ideal agreement line across the entire QFI range. In contrast to the linear and polynomial kernels, the RBF kernel efficiently captures highly nonlinear and nonpolynomial correlations among the collective observables, leading to a substantial reduction of the prediction error.

These results demonstrate that the QFI is encoded in a highly nonlinear manner within the collective observable space defined by the feature set of Eq.~\eqref{setofkernel}. Despite the intrinsically nonlinear dependence of the QFI on the eigenvalues and eigenvectors of the density operator [Eq.~\eqref{qfi}], the results show that, within the low-dimensional two-qubit Hilbert space considered here, the metrological information can already be accurately predicted from low-order collective observables without explicitly requiring the full spectral decomposition of the quantum state. This observation further motivates the use of nonlinear kernel methods for extracting metrological properties directly from experimentally accessible collective measurements.

\begin{table}[ht]
\centering
\caption{
Performance comparison of different SVR kernels for predicting the quantum Fisher information from collective observables.
}
\begin{tabular}{lcccc}
\hline
\hline
Kernel
&
Train $R^2$
&
Test $R^2$
&
Test RMSE
&
Test MAE
\\
\hline

Linear
&
0.3923
&
0.3722
&
0.4899
&
0.3787
\\

Polynomial ($d=2$)
&
0.9475
&
0.9482
&
0.1458
&
0.1132
\\

RBF
&
0.9853
&
0.9819
&
0.0862
&
0.0605
\\

\hline
\hline
\end{tabular}
\label{tab1}
\end{table}

In addition, Table~\ref{tab1} presents the quantitative regression metrics obtained for the different kernel functions. Both training and test performances are reported in order to distinguish between the fitting capability of the model and its generalization performance on previously unseen quantum states. The training metrics quantify how accurately the regression model predicts the QFI within the dataset used during optimization, whereas the test metrics evaluate the predictive capability of the trained model on independent data not employed during the training stage. The simultaneous analysis of both quantities is essential for assessing whether the model captures genuine physical correlations or simply overfits the training dataset.

The quantitative metrics in Table~\ref{tab1} confirm the trend observed in Fig.~\ref{fig1}. The RBF kernel provides the highest test accuracy and lowest prediction errors. The close agreement between the training and test $R^2$ values further indicates that the learned nonlinear correlations remain stable for previously unseen quantum states, demonstrating strong generalization capability without significant overfitting.

\section{Observable Structure of Quantum Fisher Information in Two-Qubit Systems} \label{sec5}

\begin{figure}[!htbp]
\includegraphics[width=\columnwidth]{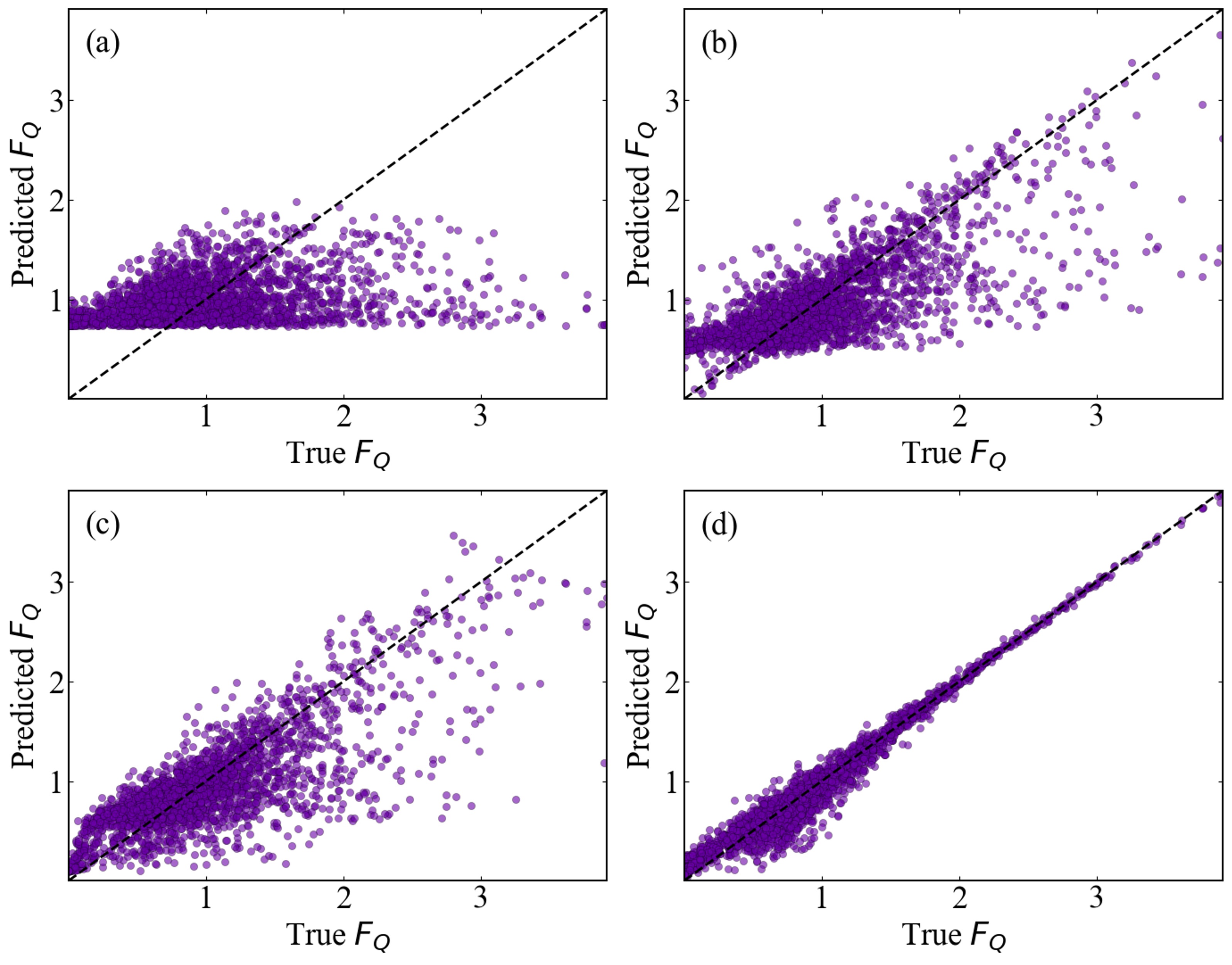} 
\caption{
Prediction of the quantum Fisher information using different sets of collective observables within the RBF-SVR model.
(a) First moments only,
(b) second moments only,
(c) symmetrized correlators only, and
(d) second moments combined with symmetrized correlators. The dashed line denotes the ideal agreement between
the exact and predicted QFI values.
}
\label{fig2}
\end{figure}

The identification of a nonlinear relationship between collective observables and the QFI naturally motivates a systematic investigation of the two-qubit prediction problem using different physically motivated feature sets. In this section, we employ the same training dataset of quantum states described previously, using the same random division of $75\%$ for training and $25\%$ for testing. All regressions were performed using the RBF kernel, while the hyperparameters $(C,\epsilon,\gamma)$ were optimized through a grid-search procedure. Figure~\ref{fig2} presents the prediction performance obtained for different classes of collective observables, where the predicted QFI values are plotted against the exact values. These results allow us to identify which sectors of the collective observable space contain the dominant metrological information associated with the QFI.

Figure~\ref{fig2}(a) shows the prediction obtained using only the first-order collective moments $\langle \hat J_i\rangle$. In this case, the predicted values exhibit large dispersion around the ideal prediction line, yielding poor predictive accuracy. This behavior indicates that average collective spin polarization alone contains very limited information about the metrological sensitivity of the quantum state to stablish a regression for QFI.

A substantial improvement is observed in Figs.~\ref{fig2}(b) and \ref{fig2}(c), where the feature sets are composed of the second-order moments $\langle \hat J_i^2\rangle$ and the symmetrized correlators $\langle \hat J_i\hat J_j+\hat J_j\hat J_i\rangle$, respectively. In both cases, the stronger concentration of points around the diagonal line demonstrates that collective fluctuations and correlations encode a significant fraction of the information governing the QFI. The comparable predictive performance of these two feature sets is consistent with the close connection between the QFI and variance-like quantities associated with the parameter-encoding generator. The highest prediction accuracy among the reduced observable sets is obtained in Fig.~\ref{fig2}(d), where second-order moments and symmetrized correlators are combined. The resulting predictions become strongly localized around the ideal agreement line, indicating that the metrological information contained in these complementary second-order observables is largely sufficient to reconstruct the QFI. 

The quantitative results reported in Table~\ref{tab2} provide further insight into the predictive power of the different observable sectors. When only the first-order moments $\langle J_i\rangle$ are employed, the regression model exhibits almost no predictive capability, yielding a test $R^2=0.0527$. In contrast, the inclusion of either the second-order moments $\langle J_i^2\rangle$ or the symmetrized correlators $\langle J_iJ_j+J_jJ_i\rangle$ increases the predictive accuracy by nearly an order of magnitude, reaching test $R^2$ values of $0.5423$ and $0.6006$, respectively. When both feature sets are combined, the prediction quality improves dramatically to $R^2=0.9656$, accompanied by a reduction of the RMSE from approximately $0.62$ to $0.12$. Although the complete feature set achieves the highest predictive accuracy, with test $R^2 \approx 0.98$, the relatively small improvement over the reduced covariance-based feature set,
$
\{\langle J_i^2\rangle,
\langle J_iJ_j+J_jJ_i\rangle\},
$
indicates that the dominant metrological information is already encoded in second-order collective observables and can be captured with remarkable accuracy using only a compact set of experimentally accessible quantities.

\begin{table}[ht]
\centering
\caption{Influence of collective observables on the prediction of the quantum Fisher information using an RBF-SVR model.}

\begin{tabular}{lcccc}
\hline
\hline

Features & Train $R^2$ & Test $R^2$ & Test RMSE & Test MAE  \\
\hline

$\langle \hat J_i\rangle$&0.0828& 0.0527 & 0.6237 & 0.4419 
\\

$\langle \hat J_i^2\rangle$ & 0.5890 & 0.5423 &  0.4336 & 0.3001 
\\

$\langle
\hat J_i\hat J_j
+
\hat J_j\hat J_i
\rangle$
&
0.5476
&
0.6006
&
0.4050
&
0.2810

\\

$\langle J_i^2\rangle,$
$\langle \hat J_i\hat J_j+\hat J_j\hat J_i\rangle$ & 0.9624 & 0.9656 & 0.1189 & 0.0841 
\\

$\langle \hat J_i\rangle,\ \langle \hat J_i\rangle^2,\ \langle \hat J_i^2\rangle,\ \langle \hat J_i\hat J_j+\hat J_j\hat J_i\rangle$ & 0.9853 & 0.9819 & 0.0862 & 0.0605 
\\

\hline
\hline
\end{tabular}
\label{tab2}
\end{table}

The learned observable-to-QFI mapping is further evaluated on five external families of two-qubit quantum states, each containing $5000$ density matrices and characterized by the feature set $\{\langle J_i^2\rangle,\langle J_iJ_j+J_jJ_i\rangle\}$. The test ensemble includes depolarized and amplitude-damped Bell-state families constructed from the maximally entangled states $\{\ket{\Phi^{+}},\ket{\Phi^{-}}\}$, together with a family of depolarized partially entangled states defined by
\begin{equation}
\rho(\theta)
=
(1-p)
\ket{\psi_{\mathrm{pure}}}
\bra{\psi_{\mathrm{pure}}}
+
p\frac{\mathbb{I}}{4},
\end{equation}
where $\ket{\psi_{\mathrm{pure}}}
=
\cos\theta\,\ket{00}
+
\sin\theta\,\ket{11}.$

The parameter $\theta$ controls the degree of entanglement of the underlying pure state, while $p$ interpolates between the pure and maximally mixed limits. Owing to the explicit dependence of Eq.~\eqref{qfi} on the spectral structure of the density operator, these families provide a stringent benchmark for assessing the robustness of the learned observable-to-QFI mapping beyond the training dataset.


\begin{table}[ht]
\centering
\caption{
Generalization performance of the trained RBF-SVR model on external two-qubit state families not included in the training dataset.
}
\begin{tabular}{lcccc}
\hline
\hline
External test family
& Test $R^2$
& Test RMSE
& Test MAE

\\
\hline

Depolarized $|\Phi^{+}\rangle$
&
0.9948
&
0.0881
&
0.0814

\\

Depolarized $|\Phi^{-}\rangle$
&
0.9939
&
0.0957
&
0.0889

\\

$\rho(\theta)$
&
0.9916
&
0.0863
&
0.0825

\\

Amplitude-damped
$|\Phi^{+}\rangle$
&
0.9966
&
0.0761
&
0.0676

\\

Amplitude-damped
$|\Phi^{-}\rangle$
&
0.9957
&
0.0856
&
0.0766

\\

\hline
\hline
\end{tabular}

\label{tab3}
\end{table}

We present the analysis of the regression model using Table~\ref{tab3}, which summarizes the quantitative performance metrics obtained for the external two-qubit state families. The model achieves consistently large test $R^2$ values, ranging from $0.9916$ to $0.9966$, demonstrating significant agreement between the predicted and exact QFI values across all external ensembles. The prediction errors remain uniformly small, with RMSE values between $0.0761$ and $0.0957$ and MAE values below $0.09$ for every state family considered. Particularly strong performance is observed for the amplitude-damped Bell states, where the model reaches $R^2=0.9966$ and RMSE$=0.0761$. These results provide strong evidence that the learned nonlinear mapping remains accurate for physically distinct quantum-state families and is not restricted to the statistical ensemble employed during training.

\begin{figure}[!htbp]
\includegraphics[width=0.8\columnwidth]{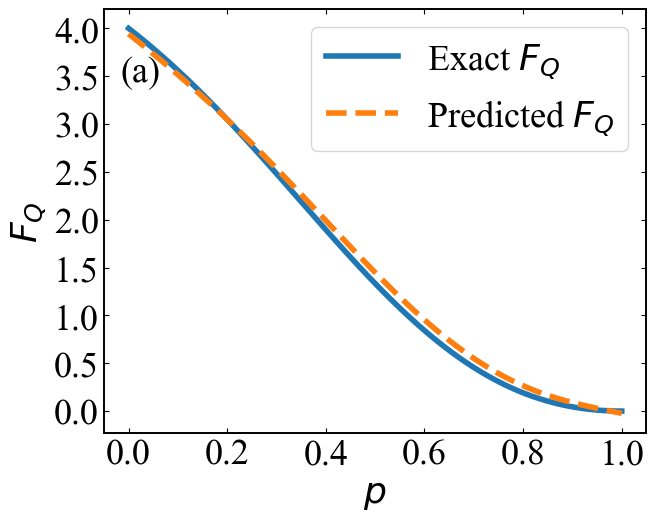} 
\caption{
Comparison between the exact and SVR-predicted quantum Fisher information for external depolarized two-qubit Bell-state family as a function of the depolarizing parameter \(p\). The solid and dashed curves represent the exact and SVR-predicted QFI values, respectively.
}
\label{fig4}
\end{figure}

We also present in Figure~\ref{fig4} the performance of the trained RBF-SVR model for depolarized two-qubit Bell states as a function of the depolarizing parameter $p$. For both Bell-state families, the predicted QFI closely follows the exact behavior across the entire noise range, reproducing the progressive loss of metrological sensitivity induced by depolarizing noise. Small deviations become visible in the strongly mixed regime ($p\rightarrow1$), indicating that part of the spectral information governing the QFI is not fully encoded within the current feature set. Nevertheless, the overall agreement demonstrates that the learned nonlinear mapping remains robust when applied to physically distinct two-qubit quantum-state families characterized by different entanglement, coherence, and spectral properties beyond those present in the training dataset.

\section{Multipartite Scaling of Quantum Fisher Information Encoding} \label{sec6}

In this section, we extend the SVR prediction framework for QFI developed for the two-qubit system in Sec.~\ref{sec5} to multipartite quantum systems composed of three, four, and five qubits. The primary objective is to investigate how the predictive capability of collective observables evolves as the Hilbert-space dimension increases and the structure of the QFI becomes progressively more complex.

The training dataset was constructed from a diverse ensemble of multipartite quantum states. Pure states of the form
$\ket{\psi}
=
\sum_{i=1}^{2^N}
c_i
\ket{i_1\cdots i_N},$,
where the complex coefficients satisfy the normalization condition
$\sum_{i=1}^{2^N}|c_i|^2=1$, mixed states were generated as convex combinations of random pure states,
$\rho_{\mathrm{mixed}}
=
\sum_{k=1}^{n}
p_k
\ket{\psi_k}\bra{\psi_k}$,
with
$\sum_{k=1}^{n}p_k=1$
and
$p_k\ge0$, and  hybrid states of the form
$\rho_{\mathrm{hybrid}}
=
t\,\rho_{\mathrm{diag}}
+
(1-t)\ket{\psi}\bra{\psi}$,
where
$0\le t\le1$
and
$\rho_{\mathrm{diag}}
=
\mathrm{diag}
(\lambda_1,\lambda_2,\ldots,\lambda_{2^N})$,
with
$\sum_{i=1}^{2^N}\lambda_i=1$
and
$\lambda_i\ge0$, together with an additional class of random quantum states generated according to the Hilbert-Schmidt construction
\begin{equation}
    \rho
    =
    \frac{AA^{\dagger}}
    {\Tr(AA^{\dagger})},
    \label{general}
\end{equation}
where $A$ denotes a random complex matrix of dimension $2^N\times 2^N$. This construction guarantees that the resulting density operator is Hermitian, positive semidefinite, and properly normalized, thereby representing a physically valid quantum state.

The complete multipartite dataset consists of $4000$ random pure states, $6000$ mixed states, $4000$ hybrid states, and $6000$ random density matrices generated through Eq.~\eqref{general}. Such a construction provides quantum states with broadly varying coherence structures, spectral properties, and state purities, allowing the regression model to explore a large region of the multipartite Hilbert space beyond highly symmetric or parametrized state families.

\subsection{Collective Spin Moment Features}

At this stage, we investigate the generalization capability of the SVR model using only collective spin-moment features, extending the analysis previously developed for the two-qubit regime. In the two-qubit case, it was shown that a relatively small set of collective spin expectation values already contains sufficient information to accurately predict the QFI.

To examine how this observable-based description scales in multipartite systems, we consider an initial 15-feature set defined by
\begin{equation}
\left\{
\langle \hat J_i\rangle,\,
\langle \hat J_i\rangle^2,\,
\langle \hat J_i^2\rangle,\,
\langle \hat J_i\hat J_j+\hat J_j\hat J_i\rangle,\,
\langle \hat J_i\rangle\langle \hat J_j\rangle
\right\},
\label{eqs15}
\end{equation}
and analyze the extent to which these collective moments remain capable of predicting the QFI for higher-dimensional multipartite quantum states. In particular, this analysis allows us to investigate whether low-order collective fluctuations and correlations continue to capture the dominant metrological information as the size of the Hilbert space increases.

\begin{figure}[!htbp]
\includegraphics[width=1.02\columnwidth]{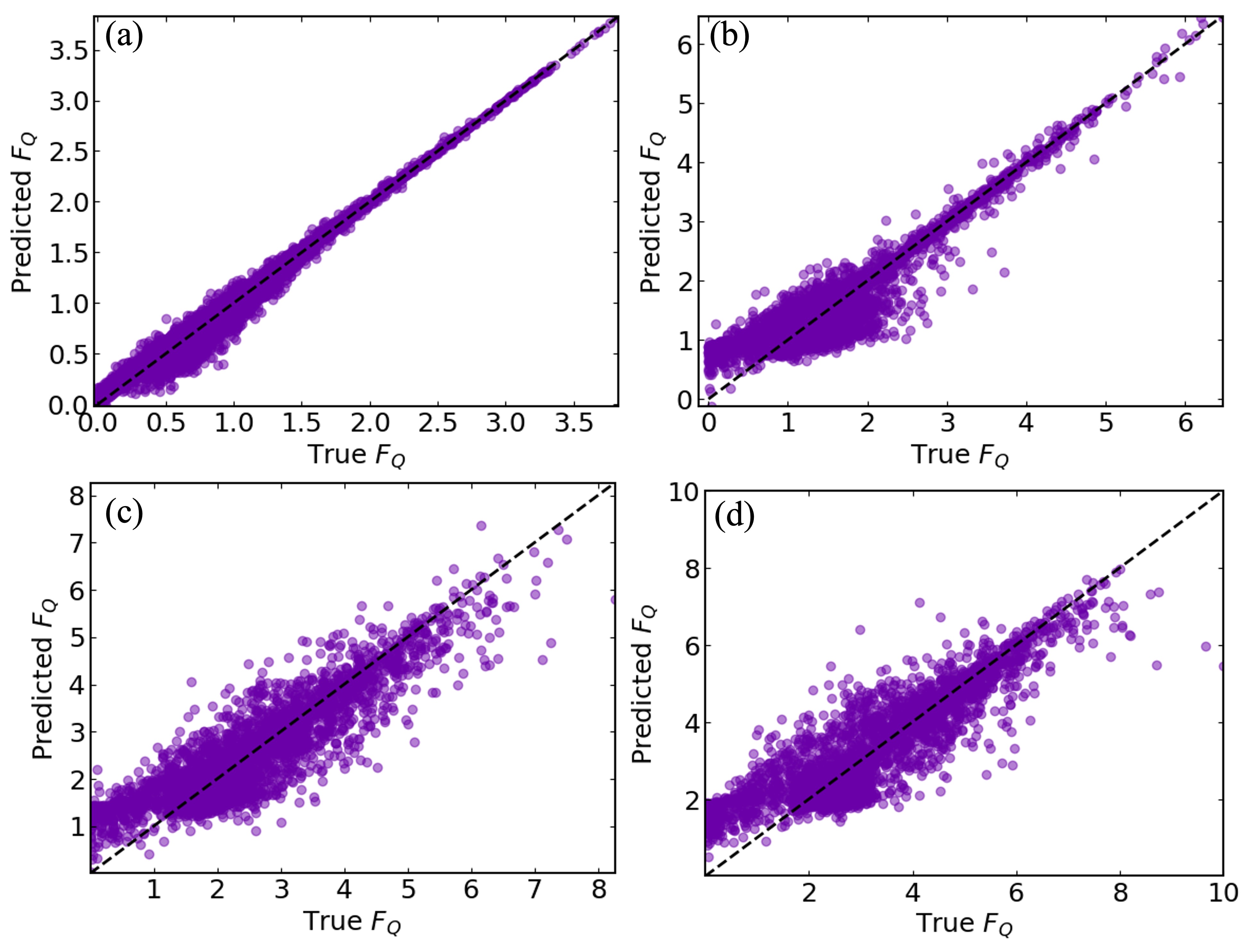} 
\caption{
Comparison between exact and SVR-predicted quantum Fisher information for multipartite systems composed of (a) two qubits, (b) three qubits, (c) four qubits, and (d) five qubits using collective spin moment features. The dashed line denotes the ideal agreement between
the exact and predicted QFI values.
}
\label{fig5}
\end{figure}

Figure~\ref{fig5} shows the dispersion plots between the predicted and exact QFI values for the two-, three-, four-, and five-qubit cases. A clear progressive degradation in the prediction accuracy is observed as the number of qubits increases, indicating that the selected 15 features set [Eq.~\eqref{eqs15}] become progressively less informative as the dimension of the Hilbert space grows. Specifically, in Fig.~\ref{fig5}(a), corresponding to the two-qubit case, the predicted values remain strongly localized around the diagonal prediction line, while
 the three-qubit system shown in Fig.~\ref{fig5}(b), noticeable deviations from the ideal prediction line begin to emerge. 
This tendency becomes more pronounced in the four- and five-qubit cases displayed in Figs.~\ref{fig5}(c) and \ref{fig5}(d). 
This behavior reflects the increasing complexity of the nonlinear structure underlying the QFI in multipartite systems, where the metrological sensitivity depends on increasingly rich many-body correlations and higher-order spectral properties of the density operator.


Nevertheless, despite the reduction in predictive accuracy for larger qubit numbers, the regression model still preserves a nontrivial global correlation between the predicted and exact QFI values across the entire metrological range. This observation indicates that collective spin moments continue to retain a substantial fraction of the physically relevant information associated with the QFI, even in multipartite regimes where the observable-to-QFI mapping acquires a significantly more intricate nonlinear structure.

\begin{table}[ht]
\centering
\caption{
Performance of the SVR model for QFI prediction using collective spin moment features.
}
\begin{tabular}{lcccc}
\hline
\hline
Qubits
&
Train $R^2$
&
Test $R^2$
&
Test RMSE
&
Test MAE
\\
\hline

2
&
0.9803
&
0.9793
&
0.0879
&
0.0612
\\

3
&
0.8651
&
0.8673
&
0.3167
&
0.2248
\\

4
&
0.8391
&
0.7901
&
0.5070
&
0.3697
\\

5
&
0.8154
&
0.7951
&
0.6097
&
0.4340
\\

\hline
\hline
\end{tabular}
\label{tab4}
\end{table}

To further quantify the trends observed in Fig.~\ref{fig5}, Table~\ref{tab4} summarizes the regression performance for multipartite systems. For two qubits, the model achieves excellent predictive accuracy, with training and test $R^2$ values of $0.9803$ and $0.9793$, respectively, together with RMSE of $0.0879$ and MAE of $0.0612$. As the Hilbert-space dimension increases, the prediction quality gradually deteriorates, with the test $R^2$ decreasing from $0.9793$ for two qubits to approximately $0.79$ for both the four- and five-qubit cases. This reduction is accompanied by a systematic increase in the prediction errors, with the RMSE growing from $0.0879$ to $0.6097$ and the MAE from $0.0612$ to $0.4340$. Nevertheless, the persistence of test $R^2$ values close to $0.8$ indicates that collective-spin observables continue to retain a significant fraction of the information governing the QFI, even though their predictive power becomes progressively limited in larger Hilbert spaces.

\subsection{Collective Spin Moment and Purity Features}

As observed in the previous subsection, collective spin moments alone become insufficient to accurately predict the QFI as the dimension of the Hilbert space increases in a SVR model. In order to enrich the physical information contained in the training dataset, we now extend the feature space by incorporating the purity of the quantum state, defined as
\(
\mathrm{Tr}(\rho^2).
\)

The inclusion of purity constitutes a physically motivated extension of the observable set, since the structure of the QFI in Eq.~\eqref{qfi} depends explicitly on the spectral properties of the density operator through its eigenvalues $\lambda_i$. While collective spin moments primarily encode information associated with fluctuations and correlations of collective observables, the purity provides additional global information regarding the degree of mixedness and coherence of the quantum state. Consequently, this quantity introduces relevant spectral information that is not fully captured by low-order collective observables alone.

In multipartite systems, the inclusion of the purity $\mathrm{Tr}(\rho^2)$ becomes particularly relevant because collective-spin observables alone do not fully characterize the spectral properties of the density operator that contribute to the QFI. Consequently, quantum states with similar values of $\langle J_i\rangle$, $\langle J_i^2\rangle$, and $\langle J_iJ_j+J_jJ_i\rangle$ may nevertheless exhibit different degrees of mixedness and, therefore, distinct metrological sensitivities. By incorporating purity into the regression model, the SVR framework gains access to complementary spectral information beyond collective fluctuations and correlations, thereby partially accounting for the spectral dependence of the QFI. Indeed, the purity can be expressed as
$\mathrm{Tr}(\rho^2)
=
\sum_i \rho_{ii}^2
+
\sum_{i\neq j}|\rho_{ij}|^2 ,$
showing that it contains contributions from both the population distribution and the total coherence content of the density operator. As a result, purity provides an additional quantity for distinguishing quantum states that appear similar at the level of collective observables but possess different spectral structures and metrological properties.

\begin{figure}[!htbp]
\includegraphics[width=1.02\columnwidth]{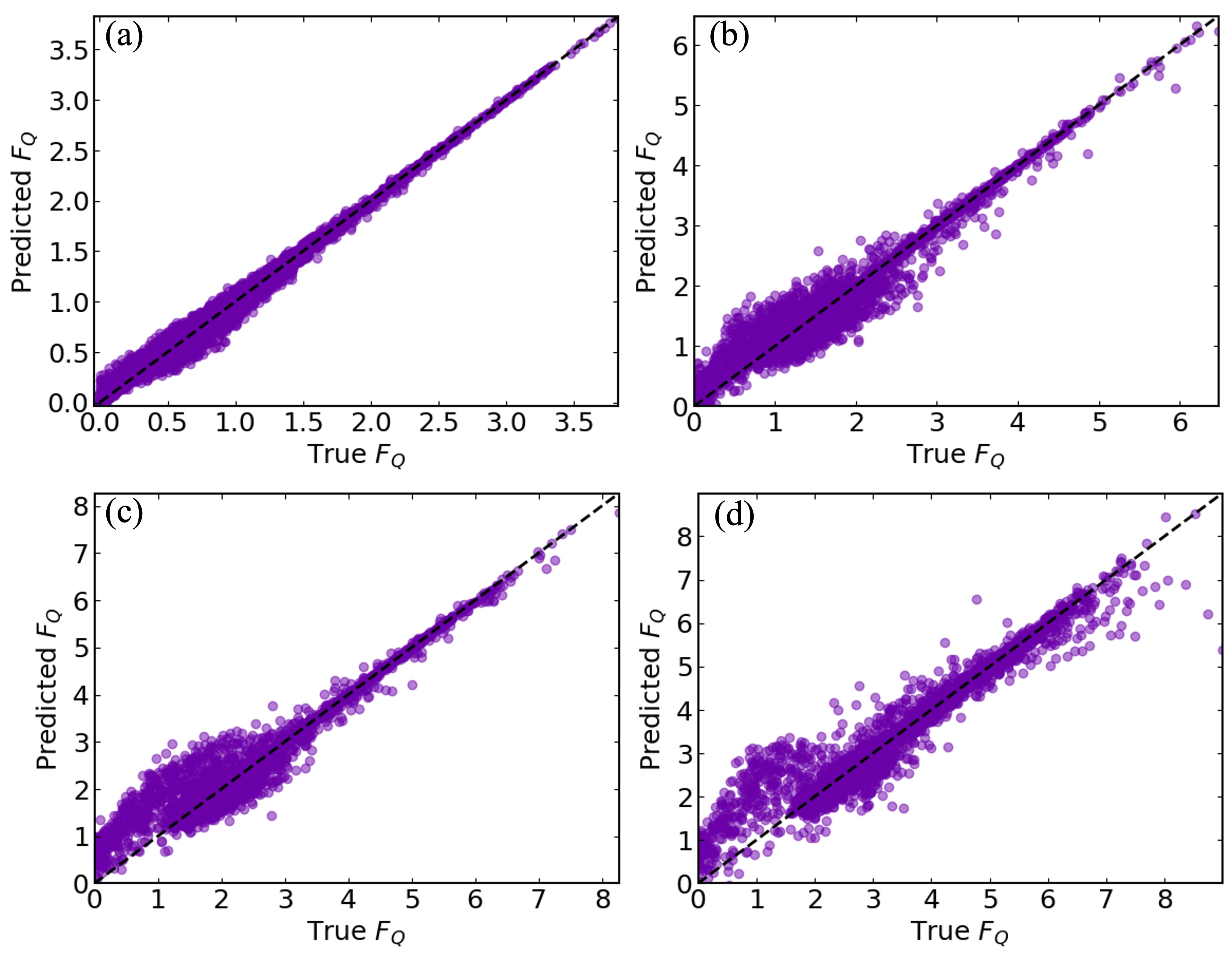} 
\caption{
Comparison between exact and SVR-predicted quantum Fisher information for multipartite systems composed of (a) two qubits, (b) three qubits, (c) four qubits, and (d) five qubits using collective spin moment and purity features. The dashed line denotes the ideal agreement between
the exact and predicted QFI values.
}
\label{fig6}
\end{figure}

Figure~\ref{fig6} shows the prediction performance after incorporating the purity as an additional feature in the regression model. Compared with Fig.~\ref{fig5}, the prediction accuracy improves substantially for all qubit configurations, indicating that the purity introduces relevant spectral information about the quantum state that is not fully encoded within low-order collective observables alone. 



\begin{table}[ht]
\centering
\caption{
Performance of the SVR model for QFI prediction using collective spin moment and purity features.
}
\begin{tabular}{lcccc}
\hline
\hline
Qubits
&
Train $R^2$
&
Test $R^2$
&
Test RMSE
&
Test MAE
\\
\hline

2
&
0.9875
&
0.9861
&
0.0722
&
0.0510
\\

3
&
0.9310
&
0.9261
&
0.2365
&
0.1637
\\

4
&
0.9251
&
0.9151
&
0.3225
&
0.1961
\\

5
&
0.9389
&
0.9194
&
0.3824
&
0.2142
\\

\hline
\hline
\end{tabular}
\label{tab5}
\end{table}

Table~\ref{tab5} summarizes the regression performance obtained after incorporating the purity $\mathrm{Tr}(\rho^2)$ into the feature set. A clear improvement is observed across all system sizes compared with the results of Table~\ref{tab4}. In particular, the test $R^2$ increases from $0.8673$, $0.7901$, and $0.7951$ to $0.9261$, $0.9151$, and $0.9194$ for the three-, four-, and five-qubit systems, respectively. Simultaneously, the RMSE decreases, while the MAE is reduced. As a result, the coefficient $R^2$ remains above $0.90$ for all multipartite configurations, indicating that purity contributes relevant spectral information that is not fully captured by low-order collective observables alone. 
Nevertheless, despite the substantial improvement obtained after incorporating $\Tr(\rho^2)$, noticeable deviations from perfect prediction still persist in higher-dimensional Hilbert spaces. This observation naturally suggests that additional higher-order spectral quantities may be required to further improve the predictive capability of the regression model. In particular, a natural extension of the feature space is the inclusion of quantities such as $\Tr(\rho^3)$, which provide additional information about the spectral decomposition of the density operator and may therefore allow the SVR model to access a more complete description of the nonlinear structure underlying the QFI.

\begin{figure}[!htbp]
\includegraphics[width=1.02\columnwidth]{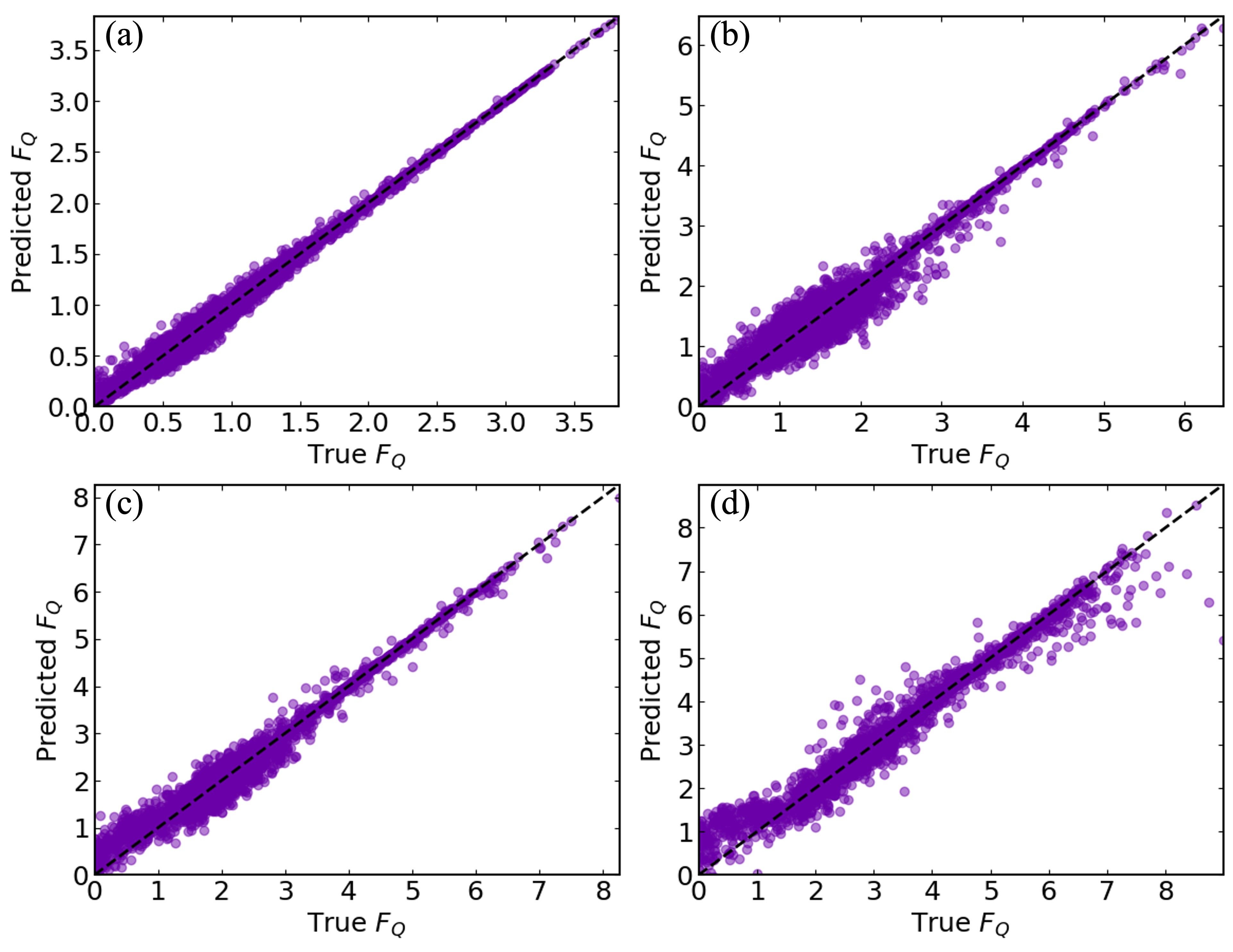} 
\caption{
Comparison between exact and SVR-predicted quantum Fisher information for multipartite systems composed of (a) two qubits, (b) three qubits, (c) four qubits, and (d) five qubits using collective spin moment, purity and Higher-Order Spectral Moment $\Tr(\rho^3)$ features. The dashed line denotes the ideal agreement between
the exact and predicted QFI values.
}
\label{fig7}
\end{figure}

\subsection{Collective Spin Moment, Purity, and Higher-Order Spectral Moment Features}

We now investigate whether the predictive performance can be further enhanced by incorporating higher-order spectral information into the feature set. 
The third spectral moment $\mathrm{Tr}(\rho^3)$ adds a further layer of spectral information. If $\rho$ has eigenvalues ${\lambda_i}$, then $\mathrm{Tr}(\rho^3)
=
\sum_i \lambda_i^3 .$
Therefore, $\mathrm{Tr}(\rho^3)$ helps distinguish states with similar purity but different eigenvalue distributions.  Thus, $\mathrm{Tr}(\rho^3)$ provides spectral information that is not contained in the purity alone. From a metrological perspective, such differences can be important because the QFI depends on eigenvalue contrasts as well as on the matrix elements of the encoding generator between the corresponding eigenvectors. This additional spectral information becomes especially important in larger Hilbert spaces, where many states can share similar low-order collective observables while possessing different metrological responses.


Figure~\ref{fig7} shows the prediction performance after simultaneously incorporating the purity and the higher-order spectral moment $\Tr(\rho^3)$ into the regression model. Compared with the previous feature configurations presented in Figs.~\ref{fig5} and \ref{fig6}, the inclusion of $\Tr(\rho^3)$ produces a further enhancement in the prediction accuracy, particularly for larger multipartite systems. This behavior indicates that higher-order spectral descriptors contain physically relevant information associated with the nonlinear structure of the QFI beyond that encoded by low-order collective observables alone.

\begin{table}[ht]
\centering
\caption{
Performance of the SVR model for QFI prediction using collective spin moment, purity and Higher-Order Spectral Moment features.
}
\begin{tabular}{lcccc}
\hline
\hline
Qubits
&
Train $R^2$
&
Test $R^2$
&
Test RMSE
&
Test MAE
\\
\hline

2
&
0.9915
&
0.9901
&
0.0607
&
0.0403
\\

3
&
0.9505
&
0.9476
&
0.1992
&
0.1389
\\

4
&
0.9721
&
0.9680
&
0.1980
&
0.1354
\\

5
&
0.9808
&
0.9644
&
0.2541
&
0.1485
\\

\hline
\hline
\end{tabular}
\label{tab6}
\end{table}

As complementary information for the prediction performance of the regression models, Table~\ref{tab6} summarizes the corresponding quantitative metrics obtained after incorporating collective spin observables together with the spectral quantities $\Tr(\rho^2)$ and $\Tr(\rho^3)$. In contrast with the previous feature configurations, the coefficient $R^2$ now remains above $0.94$ for all multipartite cases, demonstrating that the inclusion of higher-order spectral information substantially strengthens the nonlinear correlation established between the feature space and the QFI.



\subsection{Observable Structure of Quantum Fisher Information}

The predictive performance reported thus far naturally motivates a feature-level analysis of the information governing the QFI. Table~\ref{tab7} summarizes the $R^2$ coefficients obtained when the SVR model is trained using individual physically motivated feature sets for the multipartite systems considered, allowing the relative importance of different observable and spectral information sectors to be quantified.

The first moments $\langle \hat J_i \rangle$ exhibit the weakest predictive capability in the two-qubit regime, yielding an $R^2$ close to zero. This result indicates that average collective-spin polarizations alone contain very limited information about the metrological sensitivity of low-dimensional quantum states. Interestingly, the predictive power of the first moments increases progressively with the number of qubits, reaching $R^2 \approx 0.49$ for five qubits. This behavior suggests that collective polarization observables become increasingly informative in larger systems for metrological quantities, where a broader variety of collective-spin configurations can be realized.

\begin{table}[ht]
\centering
\caption{Test-set $R^2$ scores obtained for individual physically motivated feature blocks in the SVR prediction of quantum Fisher information.}
\begin{tabular}{l c c c c}
\hline
Feature set & 2 qubits & 3 qubits & 4 qubits & 5 qubits \\
\hline
$\langle \hat J_i\rangle$ & 0.1049 & 0.2537 & 0.3916 & 0.4925 \\


$\langle \hat J_i^2\rangle$ & 0.5769 & 0.5731 & 0.4726 & 0.6490 \\

$\langle \hat J_i\rangle\langle \hat J_j\rangle$ & 0.0835 & 0.1965 & 0.2145 & 0.1883 \\

$\langle \hat J_i\hat J_j+\hat J_j\hat J_i\rangle$ & 0.5824 & 0.4521 & 0.5132 & 0.5869 \\

\hline
\end{tabular}
\label{tab7}
\end{table}
\begin{table*}[ht]
\centering
\caption{Test-set $R^2$ scores obtained for combined physically motivated feature sets in the SVR prediction of quantum Fisher information.}
\begin{tabular}{l c c c c}
\hline
\hline
Feature set & 2 qubits & 3 qubits & 4 qubits & 5 qubits \\
\hline

$\langle \hat J_i^2\rangle,\langle \hat J_i \hat J_j+\hat J_j\hat J_i\rangle$ 
& 0.9640 & 0.8080 & 0.7790 & 0.7833 \\

$\langle \hat J_i\rangle^2,\langle \hat J_i\hat J_j+\hat J_j\hat J_i\rangle,\langle \hat J_i\rangle\langle \hat J_j\rangle$
& 0.7202 & 0.5326 & 0.5853 & 0.6490 \\

$\langle \hat J_i^2\rangle,\langle \hat J_i\hat J_j+\hat J_j\hat J_i\rangle,\mathrm{Tr}(\rho^2)$ 
& 0.9798 & 0.8741 & 0.9322 & 0.9349 \\

$\langle \hat J_i^2\rangle,\langle \hat J_i\hat J_j+\hat J_j\hat J_i\rangle,\mathrm{Tr}(\rho^2),\mathrm{Tr}(\rho^3)$ 
& 0.9849 & 0.8894 & 0.9447 & 0.9695 \\

$\langle \hat J_i\rangle,\langle \hat J_i^2\rangle,\langle \hat J_i\hat J_j+\hat J_j\hat J_i\rangle,\mathrm{Tr}(\rho^2),\mathrm{Tr}(\rho^3)$ 
& 0.9903 & 0.9493 & 0.9675 & 0.9778 \\

$\langle \hat J_i\rangle,\langle \hat J_i\rangle^2,\langle \hat J_i^2\rangle,\langle \hat J_i\rangle\langle \hat J_j\rangle,\langle \hat J_i\hat J_j+\hat J_j\hat J_i\rangle,\mathrm{Tr}(\rho^2),\mathrm{Tr}(\rho^3)$ 
& 0.9915 & 0.9505 & 0.9721 & 0.9808 \\

\hline
\hline
\end{tabular}
\label{tab8}
\end{table*}

A substantially stronger correlation is obtained from the second moments $\langle \hat J_i^2 \rangle$, which consistently produce $R^2$ values in the range of approximately $0.47$--$0.65$. The prominent role of these observables is physically expected, since the QFI is fundamentally related to fluctuations of the generator. In particular, for pure states the QFI reduces directly to four times the variance, making second-order moments natural carriers of metrological information. The relatively stable performance of this feature block across all qubit numbers indicates that collective fluctuations remain one of the dominant ingredients governing the QFI.

By contrast, the products of first moments, $\langle \hat J_i \rangle \langle \hat J_j \rangle$, display only limited predictive capability, yielding the smallest $R^2$ values among all considered feature sets. This result suggests that simple products of average spin polarizations do not capture the relevant fluctuation and correlation structure required to characterize the metrological response of the quantum state.

Another stronger individual performance is obtained from the symmetrized correlators,
$\langle \hat J_i \hat J_j + \hat J_j \hat J_i \rangle$, which achieve $R^2$ values comparable to, and in some cases slightly exceeding, those obtained from the second moments. These observables encode collective correlations between different spin components and therefore provide direct information about the covariance structure of the state. Their consistently large predictive power demonstrates that correlations play a central role in determining the QFI.

In conjunction, the results of Table~\ref{tab7} reveal that the dominant information governing the QFI is primarily encoded in second-order fluctuations and collective correlations rather than in average spin polarizations. This conclusion is fully consistent with the previous regression analyses and further supports the view that covariance-related quantities constitute the most informative observable sector for learning the observable-to-QFI mapping.

 Table~\ref{tab8} provides further insight into how different physically motivated feature combinations contribute to the prediction of the QFI. Unlike Table~\ref{tab7}, where each observable sector was analyzed independently, the results reported here reveal how distinct classes of observables cooperate to encode the metrological properties of multipartite quantum states.

The combination of second moments and symmetrized correlators,
$\{\langle \hat J_i^2\rangle,\langle \hat J_i\hat J_j+\hat J_j\hat J_i\rangle\}$,
already achieves significant performance in the two-qubit regime, yielding $R^2 \approx 0.96$. However, the predictive accuracy decreases for larger systems, stabilizing around $R^2\approx0.78$ for four and five qubits. 
The feature set composed of quadratic first moments and pairwise products,
$\{\langle \hat J_i\rangle^2,\langle \hat J_i\hat J_j+\hat J_j\hat J_i\rangle,\langle \hat J_i\rangle\langle \hat J_j\rangle\}$,
exhibits an intermediate predictive performance across all qubit configurations. While the obtained $R^2$ values are consistently larger than those obtained from first moments alone, they remain noticeably below those achieved by feature sets containing genuine second-order fluctuations. 
A substantial enhancement is observed once the purity $\Tr(\rho^2)$ is incorporated into the feature set $\{\langle \hat J_i^2\rangle,\langle \hat J_i\hat J_j+\hat J_j\hat J_i\rangle\}$. The predictive performance immediately exceeds $R^2\approx0.93$ for the four- and five-qubit systems, demonstrating that global spectral information provides an essential contribution to the observable-to-QFI mapping. 
The inclusion of the higher-order spectral quantity $\Tr(\rho^3)$ further improves the regression accuracy for all multipartite systems. Remarkably, the reduced feature set
$\{\langle \hat J_i^2\rangle,\langle \hat J_i\hat J_j+\hat J_j\hat J_i\rangle,\Tr(\rho^2),\Tr(\rho^3)\}$
already yields significant $R^2$ values for every qubit configuration considered. The relatively small difference between this reduced description and the complete feature set indicates that most of the information relevant for predicting the QFI is already contained in the combination of covariance-related observables and low-order spectral moments.

Finally, the complete 17-feature set yields the highest predictive accuracy across all system sizes. Owing to linear dependencies among the observables, however, only eleven of these features are independent, implying that an experimental implementation would require access to only this reduced set of quantities. Moreover, the improvement relative to the feature set composed of second-order collective observables and spectral moments is comparatively modest, particularly for the four- and five-qubit systems. This behavior suggests that the information governing the QFI is primarily encoded in two complementary sectors: collective fluctuations and low-order spectral properties. Once these contributions are incorporated through covariance-related observables together with the spectral moments $\mathrm{Tr}(\rho^{2})$ and $\mathrm{Tr}(\rho^{3})$, the inclusion of additional collective observables provides only marginal gains in predictive performance, indicating that QFI estimation can be achieved using a small set of physically motivated features.

This observation has important implications for the characterization of metrological resources in multipartite quantum systems. Although an arbitrary $N$-qubit density matrix is specified by $4^{N}-1$ independent parameters, the present results demonstrate that accurate prediction of the QFI can be achieved from a substantially reduced set of physically motivated observables. For example, full state tomography requires the determination of 63, 255, and 1023 independent parameters for three-, four-, and five-qubit systems, respectively, whereas the reduced covariance-plus-spectral feature set contains only a small number of experimentally accessible quantities. Remarkably, this compact description already reproduces the QFI with high accuracy, indicating that the metrological information relevant for parameter estimation is highly concentrated within a restricted observable sector. 


\subsection{Experimental accessibility of spectral moments} \label{sec7}
 
The improvement obtained by including $\mathrm{Tr}(\rho^2)$ and $\mathrm{Tr}(\rho^3)$ indicates that the QFI is not fully determined by low-order collective spin moments alone. This is physically expected, since the QFI depends explicitly on the spectral structure of the density matrix, including both its eigenvalues and eigenvectors. The collective spin quantities in Eq.~\eqref{eqs15} are expectation values of Hermitian operators and can therefore be directly accessed through appropriate measurements. By contrast, $\mathrm{Tr}(\rho^2)$ and $\mathrm{Tr}(\rho^3)$ are nonlinear functionals of the density matrix rather than expectation values of single-copy observables. Nevertheless, these spectral moments can be accessed without complete quantum-state tomography, for example through multi-copy overlap measurements \cite{ekert2002direct}. 

\begin{figure}[!htbp]
\centering
\includegraphics[width=0.9\columnwidth]{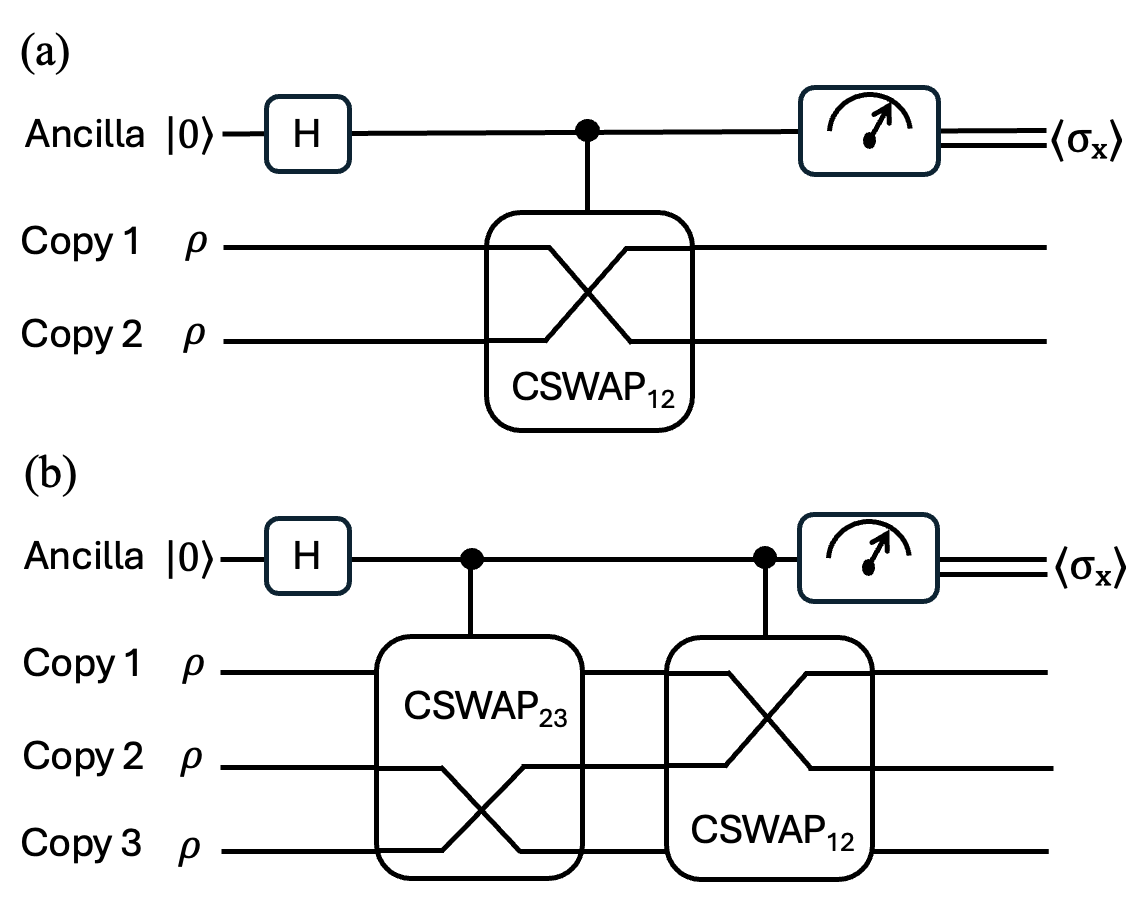}
\caption{
Interferometric circuits for measuring spectral moments.
(a) Two-copy controlled-SWAP measurement of the purity $\mathrm{Tr}(\rho^2)$. An ancilla is prepared in $|0\rangle$, transformed by a Hadamard gate, and used to control a SWAP operation between two identical copies of $\rho$. Measuring $\sigma_x$ on the ancilla gives $\langle\sigma_x\rangle=\mathrm{Tr}(\rho^2)$.
(b) Three-copy cyclic-permutation measurement of $\mathrm{Tr}(\rho^3)$. The cyclic operation $V_3=S_{12}S_{23}$ is implemented using two controlled-SWAP operations. The ancilla readout gives $\langle\sigma_x\rangle=\mathrm{Tr}(\rho^3)$.
}
\label{circuits}
\end{figure}

Experimentally, the purity can be measured using a controlled-SWAP interferometer [see Fig. \ref{circuits}(a)]. One prepares two identical copies of the state, $\rho^{\otimes 2}$, together with an ancilla qubit initialized in $|0\rangle$. After applying a Hadamard gate to the ancilla, $|0\rangle\rightarrow |+\rangle=(|0\rangle+|1\rangle)/\sqrt{2}$, one applies the controlled-SWAP (CSWAP) operation
\begin{equation}
CSWAP_{12}
=
|0\rangle\langle 0|\otimes I
+
|1\rangle\langle 1|\otimes S_{12}.
\end{equation}
The purity then follows from the standard two-copy SWAP identity $\mathrm{Tr}\left[S_{12}\rho^{\otimes 2}\right]
=
\mathrm{Tr}(\rho^2),$
where $S_{12}$ swaps the two copies of the state \cite{ekert2002direct}. In the interferometric implementation, measuring $\sigma_x$ on the ancilla gives $\langle \sigma_x\rangle=\mathrm{Tr}(\rho^2)$.

Similarly, the third moment can be obtained from the three-copy cyclic permutation operator $V_3$, defined by $V_3|\psi_1,\psi_2,\psi_3\rangle
=
|\psi_3,\psi_1,\psi_2\rangle .$ So,
for three identical copies of the state, this operator satisfies $\mathrm{Tr}\left[V_3\rho^{\otimes 3}\right]
=
\mathrm{Tr}(\rho^3)$ \cite{ekert2002direct}.
And, the corresponding controlled cyclic operation is
\begin{equation}
C(V_3)
=
|0\rangle\langle 0|\otimes I
+
|1\rangle\langle 1|\otimes V_3.
\end{equation}
The cyclic permutation can be decomposed into two ordinary SWAP operations, $V_3
=
S_{12}S_{23}.$
Accordingly, the controlled cyclic operation can be implemented as $C(V_3)
=
CSWAP_{12}\ CSWAP_{23}.$
Hence, the three-copy circuit uses two controlled-SWAP operations acting on copies 2 and 3, followed by copies 1 and 2. Measuring $\sigma_x$ on the ancilla then gives $\langle \sigma_x\rangle=\mathrm{Tr}(\rho^3)$ [see Fig. \ref{circuits} (b)].

These identities show that the spectral moments used in the regression model have a direct operational meaning. Collective spin moments encode fluctuation and correlation information, while $\mathrm{Tr}(\rho^2)$ and $\mathrm{Tr}(\rho^3)$ encode global mixedness and higher-order spectral structure. Their combination therefore provides a physically motivated and experimentally meaningful feature hierarchy for estimating QFI beyond complete quantum-state tomography.

\section{conclusion} \label{sec8}

In this paper, we developed a support vector regression framework for predicting the quantum Fisher information from physically motivated observables and low-order spectral properties of quantum states. We created large training datasets composed of random quantum states containing between two and five qubits and systematically analyzed the predictive power of collective-spin observables, low-order spectral moments, and their combinations. Our results reveal a strongly nonlinear relationship between the QFI and experimentally accessible quantities, for which the RBF kernel yields the highest predictive accuracy. In the two-qubit regime, we found that second-order collective moments and symmetrized correlations already encode most of the relevant metrological information, allowing the QFI to be predicted with high accuracy using only a compact set of collective observables.
As the Hilbert-space dimension increases, however, the predictive capability of collective observables alone progressively deteriorates, revealing that the sensitive information governing the QFI cannot be fully captured by low-order moments and correlations. We showed that this loss of information is largely recovered through the incorporation of spectral quantities, particularly the purity $\mathrm{Tr}(\rho^2)$ and the cubic spectral moment $\mathrm{Tr}(\rho^3)$. These results reveal a complementary interplay between collective observables and spectral properties in determining the metrological sensitivity of multipartite quantum states. Beyond providing an efficient machine-learning framework for estimating the quantum Fisher information, our findings reveal the observable and spectral information sectors that contribute most significantly to its determination. These results suggest that the metrological sensitivity of multipartite quantum states can be accurately characterized from a reduced set of experimentally accessible quantities, alleviating the substantial experimental and computational costs associated with full quantum-state tomography. These results further reveal how the mathematical structure of the QFI is encoded in collective correlations and spectral properties of the density operator.

\section*{Data Availability}

The codes used to generate all numerical results reported in this work is publicly available in Ref.~\cite{qfisvr2026}.

\bibliography{refs}

@article{yang2026alpha,
  title={$\alpha$-decay half-lives of superheavy nuclei with support-vector regression},
  author={Yang, Haitao and Li, Xiaopan and Song, Xiefei and Ma, Dianxu and Yu, Gongming and Bao, Xiaojun},
  journal={Physical Review C},
  volume={113},
  number={1},
  pages={014307},
  year={2026},
  publisher={APS}
}

@article{lin2026machine,
  title={Machine-learning-aided direct estimation of coherence and entanglement for unknown states},
  author={Lin, Ting and Chen, Zhihua and Wu, Kai and Guo, Zhihua and Ma, Zhihao and Fei, Shao-Ming},
  journal={Physical Review A},
  volume={113},
  number={1},
  pages={012413},
  year={2026},
  publisher={APS}
}

@book{vapnik2013nature,
  title={The nature of statistical learning theory},
  author={Vapnik, Vladimir},
  year={2013},
  publisher={Springer science \& business media}
}

@article{cortes1995support,
  title={Support-vector networks},
  author={Cortes, Corinna and Vapnik, Vladimir},
  journal={Machine learning},
  volume={20},
  number={3},
  pages={273--297},
  year={1995},
  publisher={Springer}
}

@book{bishop2006pattern,
  title={Pattern recognition and machine learning},
  author={Bishop, Christopher M and Nasrabadi, Nasser M},
  volume={4},
  number={4},
  year={2006},
  publisher={Springer}
}

@article{pan2025estimating,
  title={Estimating quantum discord for two-qubit quantum systems via machine learning},
  author={Pan, Guo-Zhu and Zhao, Jun-Long and Zhou, Jian and Yuan, Hao and Zhang, Gang},
  journal={Physica Scripta},
  volume={100},
  number={10},
  pages={105102},
  year={2025},
  publisher={IOP Publishing}
}

@article{hyllus2012fisher,
  title={Fisher information and multiparticle entanglement},
  author={Hyllus, Philipp and Laskowski, Wies{\l}aw and Krischek, Roland and Schwemmer, Christian and Wieczorek, Witlef and Weinfurter, Harald and Pezz{\'e}, Luca and Smerzi, Augusto},
  journal={Physical Review A, Atomic, Molecular, and Optical Physics},
  volume={85},
  number={2},
  pages={022321},
  year={2012},
  publisher={APS}
}

@Article{app14167312,
AUTHOR = {Zhang, Lin and Chen, Liang and He, Qiliang and Zhang, Yeqi},
TITLE = {Quantifying Quantum Coherence Using Machine Learning Methods},
JOURNAL = {Applied Sciences},
VOLUME = {14},
YEAR = {2024},
NUMBER = {16},
ARTICLE-NUMBER = {7312},
URL = {https://www.mdpi.com/2076-3417/14/16/7312},
ISSN = {2076-3417},
DOI = {10.3390/app14167312}
}

@article{mantilla2026measurement,
  title={Measurement-based quantum machine learning},
  author={Mantilla Calder{\'o}n, Luis and Raussendorf, Robert and Feldmann, Polina and Bondarenko, Dmytro},
  journal={Physical Review A},
  volume={113},
  number={4},
  pages={042421},
  year={2026},
  publisher={APS}
}

@article{helstrom1969quantum,
  title={Quantum detection and estimation theory},
  author={Helstrom, Carl W},
  journal={Journal of statistical physics},
  volume={1},
  number={2},
  pages={231--252},
  year={1969},
  publisher={Springer}
}

@article{gudder1985holevo,
  title={AS Holevo, Probabilistic and statistical aspects of quantum theory},
  author={Gudder, SP},
  year={1985}
}

@article{braunstein1994statistical,
  title={Statistical distance and the geometry of quantum states},
  author={Braunstein, Samuel L and Caves, Carlton M},
  journal={Physical Review Letters},
  volume={72},
  number={22},
  pages={3439},
  year={1994},
  publisher={APS}
}

@book{nielsen2010quantum,
  title={Quantum computation and quantum information},
  author={Nielsen, Michael A and Chuang, Isaac L},
  year={2010},
  publisher={Cambridge university press}
}

@article{pirandola2018advances,
  title={Advances in photonic quantum sensing},
  author={Pirandola, Stefano and Bardhan, B Roy and Gehring, Tobias and Weedbrook, Christian and Lloyd, Seth},
  journal={Nature Photonics},
  volume={12},
  number={12},
  pages={724--733},
  year={2018},
  publisher={Nature Publishing Group UK London}
}

@book{vonNeumann1955,
  author    = {John von Neumann},
  title     = {Mathematical Foundations of Quantum Mechanics},
  publisher = {Princeton University Press},
  year      = {1955},
  address   = {Princeton, NJ}
}

@article{4h4b-3xss,
  title = {Information conservation relations for weak measurement and its reversal},
  author = {Maleki, Yusef and Palma, Luis D. Zambrano and Zubairy, M. Suhail},
  journal = {Phys. Rev. A},
  volume = {113},
  issue = {1},
  pages = {012215},
  numpages = {7},
  year = {2026},
  month = {Jan},
  publisher = {American Physical Society},
  doi = {10.1103/4h4b-3xss},
  url = {https://link.aps.org/doi/10.1103/4h4b-3xss}
}

@article{doi:10.1142/S0219477525400280,
author = {Palma, Luis D. Zambrano and Maleki, Yusef and Zubairy, M. Suhail},
title = {Information-Theoretic Analysis of Weak Measurements and Their Reversal},
journal = {Fluctuation and Noise Letters},
volume = {25},
number = {02},
pages = {2540028},
year = {2026},
doi = {10.1142/S0219477525400280},

URL = { 
    
        https://doi.org/10.1142/S0219477525400280
    
    

},
eprint = { 
    
        https://doi.org/10.1142/S0219477525400280
    
    }
}

@article{palma2026optimal,
  title={Optimal Quantum Illumination with Nonlocal Non-Gaussian Operations},
  author={Palma, Luis D Zambrano and Maleki, Yusef and Zubairy, M Suhail},
  journal={arXiv preprint arXiv:2605.13747},
  year={2026}
}

@article{giovannetti2004quantum,
  title={Quantum-enhanced measurements: beating the standard quantum limit},
  author={Giovannetti, Vittorio and Lloyd, Seth and Maccone, Lorenzo},
  journal={Science},
  volume={306},
  number={5700},
  pages={1330--1336},
  year={2004},
  publisher={American Association for the Advancement of Science}
}

@article{giovannetti2006quantum,
  title={Quantum metrology},
  author={Giovannetti, Vittorio and Lloyd, Seth and Maccone, Lorenzo},
  journal={Physical review letters},
  volume={96},
  number={1},
  pages={010401},
  year={2006},
  publisher={APS}
}

@article{pezze2018quantum,
  title={Quantum metrology with nonclassical states of atomic ensembles},
  author={Pezze, Luca and Smerzi, Augusto and Oberthaler, Markus K and Schmied, Roman and Treutlein, Philipp},
  journal={Reviews of Modern Physics},
  volume={90},
  number={3},
  pages={035005},
  year={2018},
  publisher={APS}
}

@article{paris2009quantum,
  title={Quantum estimation for quantum technology},
  author={Paris, Matteo GA},
  journal={International Journal of Quantum Information},
  volume={7},
  number={supp01},
  pages={125--137},
  year={2009},
  publisher={World Scientific}
}

@article{pezze2009entanglement,
  title={Entanglement, nonlinear dynamics, and the Heisenberg limit},
  author={Pezz{\'e}, Luca and Smerzi, Augusto},
  journal={Physical review letters},
  volume={102},
  number={10},
  pages={100401},
  year={2009},
  publisher={APS}
}

@article{hauke2016measuring,
  title={Measuring multipartite entanglement through dynamic susceptibilities},
  author={Hauke, Philipp and Heyl, Markus and Tagliacozzo, Luca and Zoller, Peter},
  journal={Nature Physics},
  volume={12},
  number={8},
  pages={778--782},
  year={2016},
  publisher={Nature Publishing Group UK London}
}

@article{smerzi2012zeno,
  title={Zeno dynamics, indistinguishability of state, and entanglement},
  author={Smerzi, Augusto},
  journal={Physical Review Letters},
  volume={109},
  number={15},
  pages={150410},
  year={2012},
  publisher={APS}
}

@article{quek2021adaptive,
  title={Adaptive quantum state tomography with neural networks},
  author={Quek, Yihui and Fort, Stanislav and Ng, Hui Khoon},
  journal={npj Quantum Information},
  volume={7},
  number={1},
  pages={105},
  year={2021},
  publisher={Nature Publishing Group UK London}
}

@article{lohani2020machine,
  title={Machine learning assisted quantum state estimation},
  author={Lohani, Sanjaya and Kirby, Brian T and Brodsky, Michael and Danaci, Onur and Glasser, Ryan T},
  journal={Machine Learning: Science and Technology},
  volume={1},
  number={3},
  pages={035007},
  year={2020},
  publisher={IOP Publishing}
}

@Article{10.21468/SciPostPhys.7.1.009,
	title={{QuCumber: wavefunction reconstruction with neural networks}},
	author={Matthew J. S. Beach and Isaac De Vlugt and Anna Golubeva and Patrick Huembeli and Bohdan Kulchytskyy and Xiuzhe Luo and Roger G. Melko and Ejaaz Merali and Giacomo Torlai},
	journal={SciPost Phys.},
	volume={7},
	pages={009},
	year={2019},
	publisher={SciPost},
	doi={10.21468/SciPostPhys.7.1.009},
	url={https://scipost.org/10.21468/SciPostPhys.7.1.009},
}

@article{nautrup2019optimizing,
  title={Optimizing quantum error correction codes with reinforcement learning},
  author={Nautrup, Hendrik Poulsen and Delfosse, Nicolas and Dunjko, Vedran and Briegel, Hans J and Friis, Nicolai},
  journal={Quantum},
  volume={3},
  pages={215},
  year={2019},
  publisher={Verein zur F{\"o}rderung des Open Access Publizierens in den Quantenwissenschaften}
}

@article{ferrer2026physics,
  title={Physics-Informed Neural Networks for Maximizing Quantum Fisher Information in Time-Dependent Many-Body Systems},
  author={Ferrer-S{\'a}nchez, Antonio and Vives-Gilabert, Yolanda and Ban, Yue and Chen, Xi and Mart{\'\i}n-Guerrero, Jos{\'e} D},
  journal={arXiv preprint arXiv:2604.18506},
  year={2026}
}

@article{khoo2021quantum,
  title={Quantum entanglement recognition},
  author={Khoo, Jun Yong and Heyl, Markus},
  journal={Physical Review Research},
  volume={3},
  number={3},
  pages={033135},
  year={2021},
  publisher={APS}
}

@article{harney2020entanglement,
  title={Entanglement classification via neural network quantum states},
  author={Harney, Cillian and Pirandola, Stefano and Ferraro, Alessandro and Paternostro, Mauro},
  journal={New Journal of Physics},
  volume={22},
  number={4},
  pages={045001},
  year={2020},
  publisher={IOP Publishing}
}

@article{lu2017separability,
  title={A separability-entanglement classifier via machine learning},
  author={Lu, Sirui and Huang, Shilin and Li, Keren and Li, Jun and Chen, Jianxin and Lu, Dawei and Ji, Zhengfeng and Shen, Yi and Zhou, Duanlu and Zeng, Bei},
  journal={arXiv preprint arXiv:1705.01523},
  year={2017}
}

@article{greenwood2023machine,
  title={Machine-learning-derived entanglement witnesses},
  author={Greenwood, Alexander CB and Wu, Larry TH and Zhu, Eric Y and Kirby, Brian T and Qian, Li},
  journal={Physical Review Applied},
  volume={19},
  number={3},
  pages={034058},
  year={2023},
  publisher={APS}
}

@article{ren2019steerability,
  title={Steerability detection of an arbitrary two-qubit state via machine learning},
  author={Ren, Changliang and Chen, Changbo},
  journal={Physical Review A},
  volume={100},
  number={2},
  pages={022314},
  year={2019},
  publisher={APS}
}

@article{martinez2026entanglement,
  title={Entanglement detection with quantum-inspired kernels and svms},
  author={Mart{\'\i}nez-Sabiote, Ana and Skotiniotis, Michalis and Bermejo-Vega, Jara J and Manzano, Daniel and Cano, Carlos},
  journal={The Journal of Supercomputing},
  volume={82},
  number={2},
  pages={100},
  year={2026},
  publisher={Springer}
}

@article{seko2014machine,
  title={Machine learning with systematic density-functional theory calculations: Application to melting temperatures of single-and binary-component solids},
  author={Seko, Atsuto and Maekawa, Tomoya and Tsuda, Koji and Tanaka, Isao},
  journal={Physical Review B},
  volume={89},
  number={5},
  pages={054303},
  year={2014},
  publisher={APS}
}

@article{Feng_2024,
doi = {10.1088/1572-9494/ad4090},
url = {https://doi.org/10.1088/1572-9494/ad4090},
year = {2024},
month = {jun},
publisher = {IOP Publishing},
volume = {76},
number = {7},
pages = {075104},
author = {Feng, Changchun and Chen, Lin},
title = {Quantifying quantum entanglement via machine learning models},
journal = {Communications in Theoretical Physics},
}

@Article{proceedings2019012028,
AUTHOR = {Spagnolo, Nicolò and Lumino, Alessandro and Polino, Emanuele and Rab, Adil S. and Wiebe, Nathan and Sciarrino, Fabio},
TITLE = {Machine Learning for Quantum Metrology},
JOURNAL = {Proceedings},
VOLUME = {12},
YEAR = {2019},
NUMBER = {1},
ARTICLE-NUMBER = {28},
URL = {https://www.mdpi.com/2504-3900/12/1/28},
ISSN = {2504-3900},
DOI = {10.3390/proceedings2019012028}
}

@article{Zhang2020Machine,
  author = {Zhang, Ye-Qi and Yang, Li-Juan and He, Qi-Liang and Chen, Liang},
  title = {Machine learning on quantifying quantum steerability},
  journal = {Quantum Information Processing},
  year = {2020},
  volume = {19},
  number = {8},
  pages = {263},
  doi = {10.1007/s11128-020-02769-4},
  url = {https://doi.org/10.1007/s11128-020-02769-4},
  isbn = {1573-1332}
}

@article{vitale2024robust,
  title={Robust estimation of the quantum fisher information on a quantum processor},
  author={Vitale, Vittorio and Rath, Aniket and Jurcevic, Petar and Elben, Andreas and Branciard, Cyril and Vermersch, Beno{\^\i}t},
  journal={PRX Quantum},
  volume={5},
  number={3},
  pages={030338},
  year={2024},
  publisher={APS}
}

@misc{qfisvr2026,
  howpublished = {\url{https://github.com/Ldzambra/QFI-SVR}}
}

@article{huang2025quantum,
  title={Quantum metrology assisted by machine learning},
  author={Huang, Jiahao and Zhuang, Min and Zhou, Jungeng and Shen, Yi and Lee, Chaohong},
  journal={Advanced Quantum Technologies},
  volume={8},
  number={4},
  pages={2300329},
  year={2025},
  publisher={Wiley Online Library}
}

@article{lin2023quantifying,
  title={Quantifying quantum entanglement via a hybrid quantum-classical machine learning framework},
  author={Lin, Xiaodie and Chen, Zhenyu and Wei, Zhaohui},
  journal={Physical Review A},
  volume={107},
  number={6},
  pages={062409},
  year={2023},
  publisher={APS}
}

@article{vintskevich2023classification,
  title={Classification of four-qubit entangled states via machine learning},
  author={Vintskevich, SV and Bao, N and Nomerotski, A and Stankus, P and Grigoriev, DA},
  journal={Physical Review A},
  volume={107},
  number={3},
  pages={032421},
  year={2023},
  publisher={APS}
}

@article{belliardo2024applications,
  title={Applications of model-aware reinforcement learning in Bayesian quantum metrology},
  author={Belliardo, Federico and Zoratti, Fabio and Giovannetti, Vittorio},
  journal={Physical Review A},
  volume={109},
  number={6},
  pages={062609},
  year={2024},
  publisher={APS}
}

@article{fallani2022learning,
  title={Learning feedback control strategies for quantum metrology},
  author={Fallani, Alessio and Rossi, Matteo AC and Tamascelli, Dario and Genoni, Marco G},
  journal={PRX Quantum},
  volume={3},
  number={2},
  pages={020310},
  year={2022},
  publisher={APS}
}

@article{koutny2023deep,
  title={Deep learning of quantum entanglement from incomplete measurements},
  author={Koutn{\`y}, Dominik and Gin{\'e}s, Laia and Mocza{\l}a-Dusanowska, Magdalena and H{\"o}fling, Sven and Schneider, Christian and Predojevi{\'c}, Ana and Je{\v{z}}ek, Miroslav},
  journal={Science Advances},
  volume={9},
  number={29},
  pages={eadd7131},
  year={2023},
  publisher={American Association for the Advancement of Science}
}

@article{gao2024correlation,
  title={Correlation-pattern-based continuous variable entanglement detection through neural networks},
  author={Gao, Xiaoting and Isoard, Mathieu and Sun, Fengxiao and Lopetegui, Carlos E and Xiang, Yu and Parigi, Valentina and He, Qiongyi and Walschaers, Mattia},
  journal={Physical Review Letters},
  volume={132},
  number={22},
  pages={220202},
  year={2024},
  publisher={APS}
}

@article{PhysRevA.105.032408,
  title = {Detecting the steerability bounds of generalized Werner states via a backpropagation neural network},
  author = {Zhang, Jun and He, Kan and Zhang, Ying and Hao, Yu-yang and Hou, Jin-chuan and Lan, Fang-Peng and Niu, Bao-Ning},
  journal = {Phys. Rev. A},
  volume = {105},
  issue = {3},
  pages = {032408},
  numpages = {8},
  year = {2022},
  month = {Mar},
  publisher = {American Physical Society},
  doi = {10.1103/PhysRevA.105.032408},
  url = {https://link.aps.org/doi/10.1103/PhysRevA.105.032408}
}

@article{PhysRevA.104.052427,
  title = {Einstein-Podolsky-Rosen steering based on semisupervised machine learning},
  author = {Zhang, Lifeng and Chen, Zhihua and Fei, Shao-Ming},
  journal = {Phys. Rev. A},
  volume = {104},
  issue = {5},
  pages = {052427},
  numpages = {8},
  year = {2021},
  month = {Nov},
  publisher = {American Physical Society},
  doi = {10.1103/PhysRevA.104.052427},
  url = {https://link.aps.org/doi/10.1103/PhysRevA.104.052427}
}

@article{wang2024deep,
  title={Deep learning the hierarchy of steering measurement settings of qubit-pair states},
  author={Wang, Hong-Ming and Ku, Huan-Yu and Lin, Jie-Yien and Chen, Hong-Bin},
  journal={Communications Physics},
  volume={7},
  number={1},
  pages={72},
  year={2024},
  publisher={Nature Publishing Group UK London}
}

@article{PhysRevLett.123.190401,
  title = {Experimental Simultaneous Learning of Multiple Nonclassical Correlations},
  author = {Yang, Mu and Ren, Chang-liang and Ma, Yue-chi and Xiao, Ya and Ye, Xiang-Jun and Song, Lu-Lu and Xu, Jin-Shi and Yung, Man-Hong and Li, Chuan-Feng and Guo, Guang-Can},
  journal = {Phys. Rev. Lett.},
  volume = {123},
  issue = {19},
  pages = {190401},
  numpages = {6},
  year = {2019},
  month = {Nov},
  publisher = {American Physical Society},
  doi = {10.1103/PhysRevLett.123.190401},
  url = {https://link.aps.org/doi/10.1103/PhysRevLett.123.190401}
}

@article{PhysRevA.108.022427,
  title = {Entanglement verification with deep semisupervised machine learning},
  author = {Zhang, Lifeng and Chen, Zhihua and Fei, Shao-Ming},
  journal = {Phys. Rev. A},
  volume = {108},
  issue = {2},
  pages = {022427},
  numpages = {13},
  year = {2023},
  month = {Aug},
  publisher = {American Physical Society},
  doi = {10.1103/PhysRevA.108.022427},
  url = {https://link.aps.org/doi/10.1103/PhysRevA.108.022427}
}

@article{taghadomi2025effective,
  title={Effective detection of quantum discord by using convolutional neural networks},
  author={Taghadomi, Narjes and Mani, Azam and Fahim, Ali and Bakouei, A},
  journal={Quantum Machine Intelligence},
  volume={7},
  number={1},
  pages={40},
  year={2025},
  publisher={Springer}
}

@article{li2019machine,
  title={Machine learning study of the relationship between the geometric and entropy discord},
  author={Li, Xiao-Yu and Zhu, Qin-Sheng and Zhu, Ming-Zheng and Huang, Yi-Ming and Wu, Hao and Wu, Shao-Yi},
  journal={Europhysics Letters},
  volume={127},
  number={2},
  pages={20009},
  year={2019},
  publisher={EDP Sciences, IOP Publishing and Societ{\`a} Italiana di Fisica}
}

@article{lu2024quantum,
  title={Quantum machine learning: Classifications, challenges, and solutions},
  author={Lu, Wei and Lu, Yang and Li, Jin and Sigov, Alexander and Ratkin, Leonid and Ivanov, Leonid A},
  journal={Journal of Industrial Information Integration},
  volume={42},
  pages={100736},
  year={2024},
  publisher={Elsevier}
}

@article{luo2023detecting,
  title={Detecting genuine multipartite entanglement via machine learning},
  author={Luo, Yi-Jun and Liu, Jin-Ming and Zhang, Chengjie},
  journal={Physical Review A},
  volume={108},
  number={5},
  pages={052424},
  year={2023},
  publisher={APS}
}

@article{usenko2026continuous,
  title={Continuous-variable quantum communication},
  author={Usenko, Vladyslav C and Ac{\'\i}n, Antonio and All{\'e}aume, Romain and Andersen, Ulrik L and Diamanti, Eleni and Gehring, Tobias and Hajomer, Adnan AE and Kanitschar, Florian and Pacher, Christoph and Pirandola, Stefano and others},
  journal={Reviews of Modern Physics},
  volume={98},
  number={1},
  pages={015003},
  year={2026},
  publisher={APS}
}

@article{PhysRevLett.122.200401,
  title = {Machine Learning Nonlocal Correlations},
  author = {Canabarro, Askery and Brito, Samura\'{\i} and Chaves, Rafael},
  journal = {Phys. Rev. Lett.},
  volume = {122},
  issue = {20},
  pages = {200401},
  numpages = {6},
  year = {2019},
  month = {May},
  publisher = {American Physical Society},
  doi = {10.1103/PhysRevLett.122.200401},
  url = {https://link.aps.org/doi/10.1103/PhysRevLett.122.200401}
}

@article{maleki2024universal,
  title={Universal criteria for entanglement-assisted dynamical speed enhancement},
  author={Maleki, Yusef and Zubairy, M Suhail},
  journal={Physical Review A},
  volume={110},
  number={5},
  pages={052415},
  year={2024},
  publisher={APS}
}

@article{hasan2023quantum,
  title={Quantum communication systems: vision, protocols, applications, and challenges},
  author={Hasan, Syed Rakib and Chowdhury, Mostafa Zaman and Sayem, Mohammad and Jang, Yeong Min},
  journal={IEEE Access},
  volume={11},
  pages={15855--15877},
  year={2023},
  publisher={IEEE}
}

@article{halimeh2025cold,
  title={Cold-atom quantum simulators of gauge theories},
  author={Halimeh, Jad C and Aidelsburger, Monika and Grusdt, Fabian and Hauke, Philipp and Yang, Bing},
  journal={Nature Physics},
  volume={21},
  number={1},
  pages={25--36},
  year={2025},
  publisher={Nature Publishing Group UK London}
}

@article{daley2022practical,
  title={Practical quantum advantage in quantum simulation},
  author={Daley, Andrew J and Bloch, Immanuel and Kokail, Christian and Flannigan, Stuart and Pearson, Natalie and Troyer, Matthias and Zoller, Peter},
  journal={Nature},
  volume={607},
  number={7920},
  pages={667--676},
  year={2022},
  publisher={Nature Publishing Group UK London}
}

@article{karsa2024quantum,
  title={Quantum illumination and quantum radar: A brief overview},
  author={Karsa, Athena and Fletcher, Alasdair and Spedalieri, Gaetana and Pirandola, Stefano},
  journal={Reports on progress in physics},
  volume={87},
  number={9},
  pages={094001},
  year={2024},
  publisher={IOP Publishing}
}

@article{huang2024entanglement,
  title={Entanglement-enhanced quantum metrology: From standard quantum limit to Heisenberg limit},
  author={Huang, Jiahao and Zhuang, Min and Lee, Chaohong},
  journal={Applied Physics Reviews},
  volume={11},
  number={3},
  year={2024},
  publisher={AIP Publishing}
}

@article{maleki2022distributed,
  title={Distributed phase estimation and networked quantum sensors with W-type quantum probes},
  author={Maleki, Yusef and Zubairy, M Suhail},
  journal={Physical Review A},
  volume={105},
  number={3},
  pages={032428},
  year={2022},
  publisher={APS}
}

@article{taddei2013quantum,
  title={Quantum speed limit for physical processes},
  author={Taddei, M{\'a}rcio M and Escher, Bruno M and Davidovich, Luiz and de Matos Filho, Ruynet L},
  journal={Physical review letters},
  volume={110},
  number={5},
  pages={050402},
  year={2013},
  publisher={APS}
}

@article{wise2021using,
  title={Using deep learning to understand and mitigate the qubit noise environment},
  author={Wise, David F and Morton, John JL and Dhomkar, Siddharth},
  journal={PRX Quantum},
  volume={2},
  number={1},
  pages={010316},
  year={2021},
  publisher={APS}
}

@article{whiteson2003support,
  title={Support vector regression as a signal discriminator in high energy physics},
  author={Whiteson, Daniel O and Naumann, N Axel},
  journal={Neurocomputing},
  volume={55},
  number={1-2},
  pages={251--264},
  year={2003},
  publisher={Elsevier}
}

@article{jalili2024prediction,
  title={Prediction of ground state charge radius using support vector regression},
  author={Jalili, Amir and Chen, Ai-Xi},
  journal={New Journal of Physics},
  volume={26},
  number={10},
  pages={103017},
  year={2024},
  publisher={IOP Publishing}
}

@article{gao2009accurate,
  title={An accurate density functional theory calculation for electronic excitation energies: The least-squares support vector machine},
  author={Gao, Ting and Sun, Shi-Ling and Shi, Li-Li and Li, Hui and Li, Hong-Zhi and Su, Zhong-Min and Lu, Ying-Hua},
  journal={The Journal of chemical physics},
  volume={130},
  number={18},
  year={2009},
  publisher={AIP Publishing}
}

@article{balabin2011support,
  title={Support vector machine regression (LS-SVM)—an alternative to artificial neural networks (ANNs) for the analysis of quantum chemistry data?},
  author={Balabin, Roman M and Lomakina, Ekaterina I},
  journal={Physical Chemistry Chemical Physics},
  volume={13},
  number={24},
  pages={11710--11718},
  year={2011},
  publisher={Royal Society of Chemistry}
}

@inproceedings{vitek2013towards,
  title={Towards the modeling of atomic and molecular clusters energy by support vector regression},
  author={Vitek, Ale and Stachon, Martin and Kr{\"o}mer, Pavel and Sn{\'a}el, V{\'a}clav},
  booktitle={2013 5th International Conference on Intelligent Networking and Collaborative Systems},
  pages={121--126},
  year={2013},
  organization={IEEE}
}

@article{chowdhury2025exploring,
  title={Exploring Quantum Support Vector Regression for Predicting Hydrogen Storage Capacity of Nanoporous Materials},
  author={Chowdhury, Chandra},
  journal={Advanced Intelligent Discovery},
  pages={e202500015},
  year={2025},
  publisher={Wiley Online Library}
}

@article{varela2026entanglement,
  title={Entanglement Detection and Quantification Through Machine Learning: A Comprehensive Review},
  author={Varela, Joab M and de Palhares Jr, Alberto B and Duarte, Diogo HG},
  journal={Brazilian Journal of Physics},
  volume={56},
  number={1},
  pages={25},
  year={2026},
  publisher={Springer}
}

@article{strikis2021learning,
  title={Learning-based quantum error mitigation},
  author={Strikis, Armands and Qin, Dayue and Chen, Yanzhu and Benjamin, Simon C and Li, Ying},
  journal={PRX Quantum},
  volume={2},
  number={4},
  pages={040330},
  year={2021},
  publisher={APS}
}

@article{wallnofer2020machine,
  title={Machine learning for long-distance quantum communication},
  author={Walln{\"o}fer, Julius and Melnikov, Alexey A and D{\"u}r, Wolfgang and Briegel, Hans J},
  journal={PRX quantum},
  volume={1},
  number={1},
  pages={010301},
  year={2020},
  publisher={APS}
}

@article{chin2021machine,
  title={Machine learning aided carrier recovery in continuous-variable quantum key distribution},
  author={Chin, Hou-Man and Jain, Nitin and Zibar, Darko and Andersen, Ulrik L and Gehring, Tobias},
  journal={npj Quantum Information},
  volume={7},
  number={1},
  pages={20},
  year={2021},
  publisher={Nature Publishing Group UK London}
}

@article{pedregosa2011scikit,
  title={Scikit-learn: Machine learning in Python},
  author={Pedregosa, Fabian and Varoquaux, Ga{\"e}l and Gramfort, Alexandre and Michel, Vincent and Thirion, Bertrand and Grisel, Olivier and Blondel, Mathieu and Prettenhofer, Peter and Weiss, Ron and Dubourg, Vincent and others},
  journal={the Journal of machine Learning research},
  volume={12},
  pages={2825--2830},
  year={2011},
  publisher={JMLR. org}
}

@article{maleki2021quantum2,
  title={Quantum metrology with superposition spin coherent states: Insights from Fisher information},
  author={Maleki, Yusef and Scully, Marlan O and Zheltikov, Aleksei M},
  journal={Physical Review A},
  volume={104},
  number={5},
  pages={053712},
  year={2021},
  publisher={APS}
}

@article{vidrighin2014joint,
  title={Joint estimation of phase and phase diffusion for quantum metrology},
  author={Vidrighin, Mihai D and Donati, Gaia and Genoni, Marco G and Jin, Xian-Min and Kolthammer, W Steven and Kim, MS and Datta, Animesh and Barbieri, Marco and Walmsley, Ian A},
  journal={Nature communications},
  volume={5},
  number={1},
  pages={3532},
  year={2014},
  publisher={Nature Publishing Group UK London}
}

@article{maleki2021quantum,
  title={Quantum phase estimations with spin coherent states superposition},
  author={Maleki, Yusef},
  journal={The European Physical Journal Plus},
  volume={136},
  number={10},
  pages={1028},
  year={2021},
  publisher={Springer}
}

@article{ekert2002direct,
  title={Direct estimations of linear and nonlinear functionals of a quantum state},
  author={Ekert, Artur K and Alves, Carolina Moura and Oi, Daniel KL and Horodecki, Micha{\l} and Horodecki, Pawe{\l} and Kwek, Leong Chuan},
  journal={Physical review letters},
  volume={88},
  number={21},
  pages={217901},
  year={2002},
  publisher={APS}
}

@incollection{petz2011introduction,
  title={Introduction to quantum Fisher information},
  author={Petz, D{\'e}nes and Ghinea, Catalin},
  booktitle={Quantum probability and related topics},
  pages={261--281},
  year={2011},
  publisher={World Scientific}
}

@article{walborn2018quantum,
  title={Quantum-enhanced sensing from hyperentanglement},
  author={Walborn, SP and Pimentel, AH and Davidovich, L and de Matos Filho, RL},
  journal={Physical Review A},
  volume={97},
  number={1},
  pages={010301},
  year={2018},
  publisher={APS}
}

@article{zeng2023approximate,
  title={Approximate autonomous quantum error correction with reinforcement learning},
  author={Zeng, Yexiong and Zhou, Zheng-Yang and Rinaldi, Enrico and Gneiting, Clemens and Nori, Franco},
  journal={Physical Review Letters},
  volume={131},
  number={5},
  pages={050601},
  year={2023},
  publisher={APS}
}

@article{che2026quantum,
  title={Quantum circuit complexity and unsupervised machine learning of topological order},
  author={Che, Yanming and Gneiting, Clemens and Wang, Xiaoguang and Nori, Franco},
  journal={Nature Communications},
  year={2026},
  publisher={Nature Publishing Group UK London}
}

@article{ladd2010quantum,
  title={Quantum computers},
  author={Ladd, Thaddeus D and Jelezko, Fedor and Laflamme, Raymond and Nakamura, Yasunobu and Monroe, Christopher and O'Brien, Jeremy Lloyd},
  journal={nature},
  volume={464},
  number={7285},
  pages={45--53},
  year={2010},
  publisher={Nature Publishing Group UK London}
}

\end{document}